\documentclass[aps,prl,preprint,superscriptaddress,longbibliography]{revtex4-1}
\usepackage{upgreek}
\usepackage{amssymb, amsmath}
\usepackage{bm}
\usepackage{pifont}
\usepackage{graphicx}
\usepackage[usenames]{color}
\usepackage[colorlinks=true, pdfstartview=FitV, linkcolor=blue, citecolor=blue, urlcolor=blue]{hyperref}
\usepackage[normalem]{ulem}
\newcommand{\INH}{\sigma_{\mathrm{INH}}}
\newcommand{\BCD}{\sigma_{\mathrm{BCD}}}

\newcommand{\mathd}{\mathrm{d}}

\begin{document}

\title{Intrinsic Nonlinear Hall Detection of the N\'eel Vector for Two-Dimensional Antiferromagnetic Spintronics}

\author{Jizhang Wang}
\affiliation{School of Physics, Peking University, Beijing 100871, China}

\author{Hui Zeng}
\affiliation{State Key Laboratory of Low-Dimensional Quantum Physics, Department of Physics, Tsinghua University, Beijing 100084, China}

\author{Wenhui Duan}
\affiliation{State Key Laboratory of Low-Dimensional Quantum Physics, Department of Physics, Tsinghua University, Beijing 100084, China}
\affiliation{Institute for Advanced Study, Tsinghua University, Beijing 100084, China}
\affiliation{Frontier Science Center for Quantum Information, Beijing 100084, China}
\affiliation{Collaborative Innovation Center of Quantum Matter, Beijing 100871, China}

\author{Huaqing Huang}
\email[Corresponding author: ]{huaqing.huang@pku.edu.cn}
\affiliation{School of Physics, Peking University, Beijing 100871, China}
\affiliation{Collaborative Innovation Center of Quantum Matter, Beijing 100871, China}
\affiliation{Center for High Energy Physics, Peking University, Beijing 100871, China}

\date{\today}

\begin{abstract}
The respective unique merit of antiferromagnets and two-dimensional (2D) materials in spintronic applications inspire us to exploit 2D antiferromagnetic spintronics. However, the detection of the N\'eel vector in 2D antiferromagnets remains a great challenge because the measured signals usually decrease significantly in the 2D limit. Here we propose that the N\'eel vector of 2D antiferromagnets can be efficiently detected by the intrinsic nonlinear Hall (INH) effect which exhibits unexpected significant signals. As a specific example, we show that the INH conductivity of the monolayer manganese chalcogenides Mn$X$ ($X$=S, Se, Te) can reach the order of nm$\cdot$mA/V$^2$, which is orders of magnitude larger than experimental values of paradigmatic antiferromagnetic spintronic materials. The INH effect can be accurately controlled by shifting the chemical potential around the band edge, which is experimentally feasible via electric gating or charge doping.
Moreover, we explicitly demonstrate its $2\pi$-periodic dependence on the N\'eel vector orientation based on an effective $k.p$ model. Our findings enable flexible design schemes and promising material platforms for spintronic memory device applications based on 2D antiferromagnets.
\end{abstract}

\maketitle

\textit{Introduction.}---The desire to reduce the size and power consumption of spintronic devices stimulated the emergence of a new field referred to as two-dimensional (2D) spintronics \cite{RevModPhys.92.021003,inf2.12048,wcms.1259,linXiaoyang2019,AhnEthan2020,Hanwei2016}.
Two-dimensional materials with atomic thickness have attracted extraordinary interest in spintronics because they not only provide a promising opportunity to push the relevant devices to the 2D limit, but also enable the hopeful exploration of new spintronic phenomena
due to their unusual spin-dependent properties, such as the spin-valley coupling of transition metal dichalcogenides \cite{PhysRevLett.108.196802} and the spin-momentum locking of quantum spin Hall insulators \cite{PhysRevLett.95.226801}. However, almost all the existing 2D materials proposed for spintronics are nonmagnetic or ferromagnetic \cite{ZouXiaolong2021} (e.g., CrI$_3$ \cite{Bevin2017CrI3} and CrGeTe$_3$ \cite{Gong2017Cr2Ge2Te6}). In this Letter, we extend the 2D spintronics to antiferromagnets and show that the N\'eel vector, which serves as a state variable for 2D antiferromagnetic spintronics, can be detected by the nonlinear Hall measurement.

Antiferromagnets composed of antiferromagnetically coupled magnetic elements are attractive for spintronics because of their faster dynamics, zero stray fields, and insensitivity to magnetic perturbations \cite{Baltz.2018.Tserkovnyak,jungwirth_antiferromagnetic_2016}. The robust high-speed manipulation of the N\'eel vector \cite{PhysRevLett.113.157201, RevModPhys.91.035004}, such as ultrafast 90$^\circ$ switching by current-induced spin-orbit toque \cite{wadley_electrical_2016,olejnik_antiferromagnetic_2017, Bodnar_writingAndReadingMn2Au, PhysRevB.94.014439,Olejnik2017ThzWriting} and reproducible 180$^\circ$ reversal by flipping the polarity of the writing current \cite{wadley-CurrentPolaritydependentManipulation-2018,godinho_electrically_2018}, have been demonstrated in recent experiments. However, the failure of N\'eel vector detections via conventional magnetic techniques due to the absence of net magnetization poses a major challenge for practical applications of antiferromagnetic spintronics \cite{Baltz.2018.Tserkovnyak}.
{Several optical and microscopic methods, such as the spin-polarized scanning tunneling microscopy \cite{PhysRevLett.86.4132}, the x-ray magnetic linear dichroism microscopy \cite{PhysRevLett.118.057701}, the femtosecond pump-probe magneto-optical experiment
\cite{saidl2017optical}, the spatially resolved second-harmonic generation \cite{sun2019giant,ni2021imaging}, and the diamond nitrogen-vacancy scanning probe magnetometry %(diamond nitrogen-vacancy microscopy)
\cite{D2RA06440E} work for accurate detection of N\'eel vector, but are difficult to incorporate for high-density integration devices. The anisotropic magnetoresistance (AMR) effect is useful for experimental detection of a 90$^\circ$ rotation but is invariant upon a 180$^\circ$ reversal of the N\'eel vector, and its small magnitude limits the readout speed and the possible miniaturization \cite{zelezny2018SpinTransport}.} Recently, the reversed N\'eel vector states are electrically distinguished by a second-order magnetoresistance effect \cite{godinho_electrically_2018, PhysRevLett.127.277201, PhysRevLett.127.277202}, but has been limited in bulk materials. Since the readout speed and size scalability is usually proportional to the magnitude of the response signal which could significantly diminish in the ultimate atomic limit \cite{jungwirth_antiferromagnetic_2016, Jungwirth2018mutiple_diraction}, it is, therefore, natural to raise the question: how to efficiently detect the N\'eel vector with significantly large readout signal for 2D antiferromagnetic spintronics?

In this Letter, we predict that the N\'eel vector of 2D antiferromagnets can be efficiently read out using the INH effect which exhibits a significant signal that can be detected by experimental means. Taking 2D manganese chalcogenide Mn$X$ ($X$=S, Se, Te) as an example, we show that the INH conductivity of the MnS monolayer can reach the order of nm$\cdot$ mA/V$^2$, which is orders-of-magnitude larger than the experimentally measured values of typical antiferromagnets CuMnAs \cite{godinho_electrically_2018, PhysRevLett.127.277201} and Mn$_2$Au \cite{PhysRevLett.127.277202}. The INH effect can be controlled by shifting the chemical potential via electric gating or charge doping. We further develop an effective $k\cdot p$ model to explain its dependence on the N\'eel vector direction with a $2\pi$ periodicity.
Together with the efficient write-in approach based on current-induced spin-orbit toques, our findings constitute promising design schemes and material platforms for 2D antiferromagnetic spintronics.

\begin{figure}
\includegraphics[width =1\columnwidth]{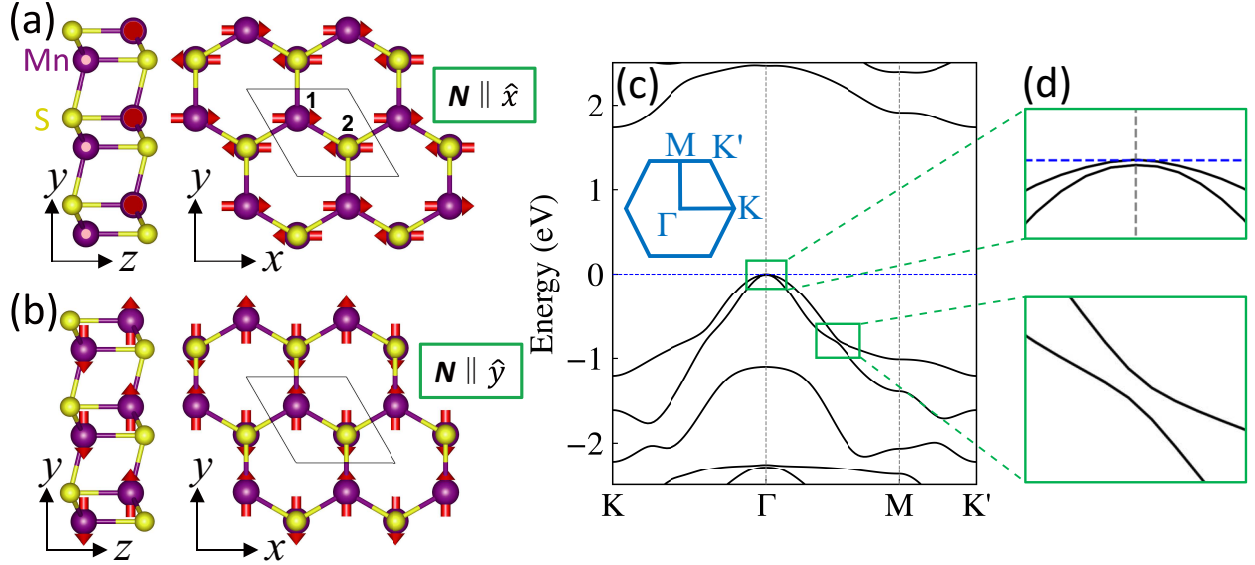}
\caption{\label{fig1} Atomic and band structure of MnS. (a),(b) The top and side view of monolayer MnS with $\bm{N}\parallel\hat{x}$ and $\bm{N}\parallel\hat{y}$. Red arrows indicate magnetic moments. (c) the band structure of MnS with $\bm{N}\parallel \hat{x}$. The insert shows the Brillouin zone. (d) The zoom-in plot of bands in small-gap regions.}
\end{figure}

\textit{Atomic and band structures.}---Due to similar crystal structures of 2D Mn$X$, which have been successfully synthesized in experiments \cite{acsnano.1c05532}, we take MnS as an example hereafter and defer the rest to the Supplemental Material \footnote{\label{fn}See Supplemental Material at http://link.aps.org/supplemental/xxx, for more details about the derivation of the effective Hamiltonian and the numerical calculation of INH effect, which include Refs.~\cite{VASP,wannier90,godinho_electrically_2018}}.
As shown in Fig.~\ref{fig1}(a), MnS crystallizes in an AA-stacked bilayer honeycomb lattice, where Mn (and $X$)
atoms on top and bottom layers (Mn$_1$ and Mn$_2$) occupy opposite sublattices. The lattice structure belongs to the space group of $P\bar{3}m1$ (No. 164, $D_{3d}^3$). Our first-principles calculations \footnotemark[\value{footnote}] show that the magnetic moments are about 4.36 $\mu_B$ per Mn atom and are antiferromagnetically ordered, which are consistent with previous studies \cite{PhysRevB.106.085410}. The N\'eel vector $\bm{N}$, which is defined as the difference of the magnetic moments between Mn$_1$ and Mn$_2$ in the unit-cell, shows a significant in-plane anisotropy with the magnetocrystalline anisotropy energy being about 0.4 meV per unit-cell. For $\bm{N}$ lying in different in-plane directions, there is little energy difference (see Fig.~S5 \footnotemark[\value{footnote}]), indicating that it is possible to electrically manipulate $\bm{N}$ by current pulses via spin-orbit torques \cite{grzybowski-ImagingCurrentInducedSwitching-2017, godinho_electrically_2018, Bodnar_writingAndReadingMn2Au,PhysRevB.99.140409}. More importantly, for an arbitrary direction of $\bm{N}$, which is denoted by its polar angle $\theta$ with respect to the $x$-axis, the combination of spatial and time reversal symmetry ($\mathcal{PT}$) is respected.

Figure~\ref{fig1}(c) shows the calculated band structure for 2D MnS with $\bm{N}\parallel\hat{x}$. Because of the $\mathcal{PT}$ symmetry, every band is doubly degenerate. It is noted that some nearly degenerate points (NDPs) lie at $\Gamma$ at the valence band maximum and along the $\Gamma-M$ line around $E\approx -0.8$ eV. Since the valence bands are dominated by the S-\textit{p} orbitals, the weak spin-orbit coupling of S only induces slight band splitting at the NDPs [see Fig.~\ref{fig1}(d)] compared to MnSe and MnTe \footnotemark[\value{footnote}], which plays a crucial role for the INH effect, as we will discuss later.

\begin{figure}
\includegraphics[width =1\columnwidth]{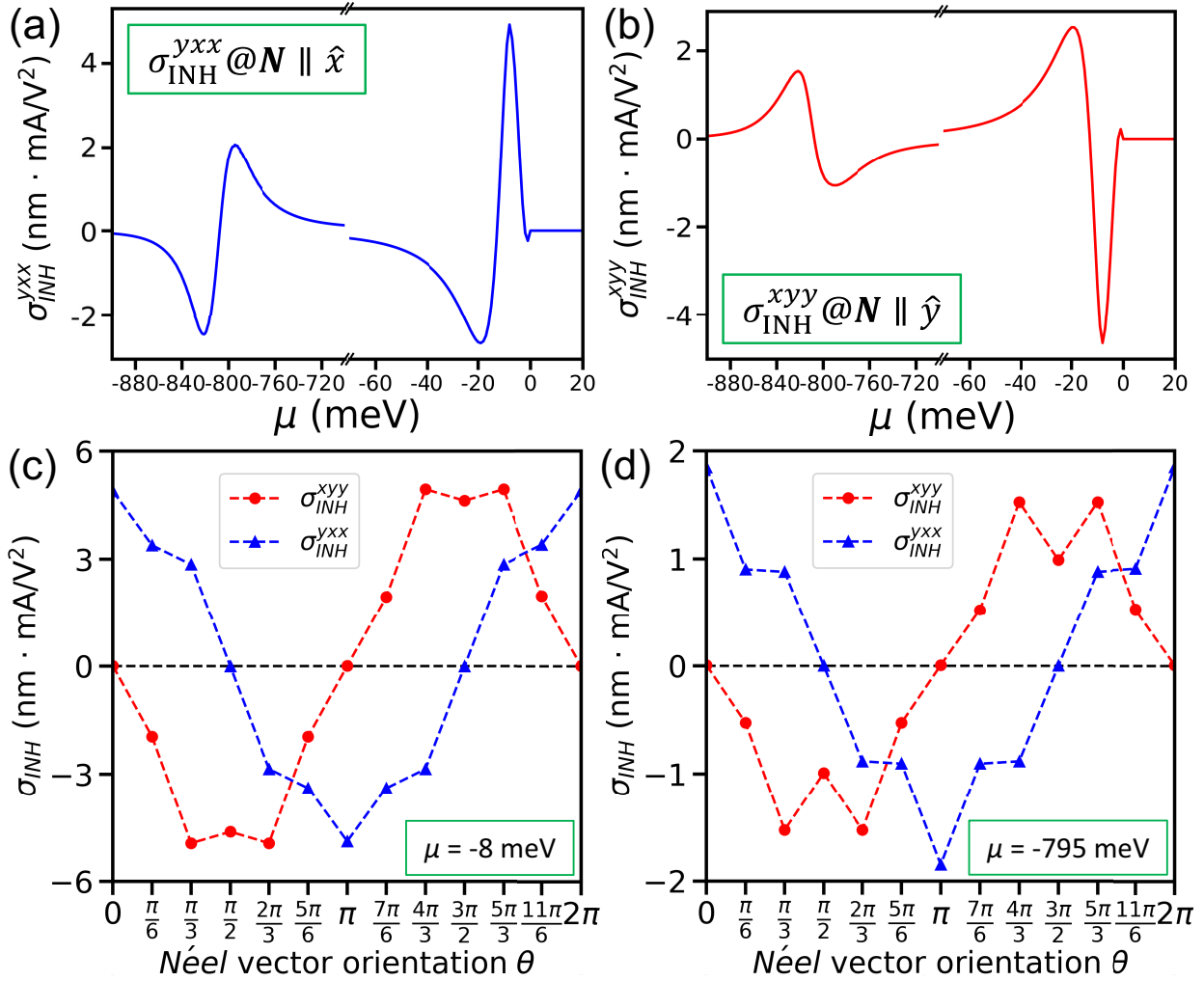}
\caption{\label{fig2} The INH conductivity of MnS. (a) $\INH^{yxx}$ for $\bm{N}\parallel\hat{x}$ and (b) $\INH^{xyy}$ for $\bm{N}\parallel\hat{y}$ as a function of the chemical potential $\mu$. (c),(d) $\INH^{xyy}$ and $\INH^{yxx}$ at (c) $\mu=-8$ meV and (d) $-795$ meV when $\bm{N}$ rotates in the $x$-$y$ plane.}
\end{figure}

\textit{INH effect in MnS.}---
In general, the nonlinear Hall conductivity tensor %$\bm{\sigma}$
is defined as the quadratic current response $\bm{J}$ to electric field $\bm{E}$: $J^\alpha=\sum_{\beta\neq\alpha,\gamma} \sigma^{\alpha\beta\gamma} E^\beta E^\gamma$, where $\alpha,\beta,\gamma$ are Cartesian indices. The nonlinear Hall conductivity can be separated into time-reversal-even ($\mathcal{T}$-even) and $\mathcal{T}$-odd parts, but only the latter can be utilized to detect the N\'eel vector reversal. In 2D antiferromagnets respecting the $\mathcal{PT}$ symmetry, the $\mathcal{T}$-even Berry curvature dipole (BCD) contribution $\BCD$ is strictly forbidden \cite{nandy2019symmetry,ma2019observation}. In contrast, the $\mathcal{T}$-odd INH conductivity $\INH$ which is allowed becomes an ideal quantity for the N\'eel vector detection, and is therefore our main concern.
The INH conductivity can be expressed in terms of band quantities as
~\cite{gao_field_2014,gao_orbital_2018}
\begin{eqnarray}
 \INH^{\alpha \beta \gamma} &=&\int_\mathrm{BZ} \frac{\mathd \bm{k}}{(2 \uppi)^d}\Lambda^{\alpha\beta\gamma}(\bm{k})\nonumber\\
 &=&\int_\mathrm{BZ} \frac{\mathd \bm{k}}{(2 \uppi)^d}\sum_n\lambda^{\alpha\beta\gamma}_n\frac{\partial f (\epsilon_n;\mu)}{\partial \epsilon_n},\label{inhexp}\\
 \lambda_n^{\alpha\beta\gamma}&=&v_n^{\alpha}G^n_{\beta\gamma}(\bm{k})-v_n^{\beta}G^n_{\alpha\gamma}(\bm{k}),\label{lambda}\\
G_{\alpha\beta}^n(\bm{k}) &=& 2e^3\mathrm{Re}\sum_{m \ne n}\frac{A_\alpha^{nm}(\bm{k})A_\beta^{mn}(\bm{k})}{\epsilon_n(\bm{k}) - \epsilon_m(\bm{k})}\label{bcp},
\end{eqnarray}
where $G_{\alpha\beta}^n(\bm{k})$ is the Berry-connection polarizability (BCP) and $\Lambda^{\alpha\beta}(\bm{k})$ [$\lambda^{\alpha\beta}_n(\bm{k})$] is the (band-resolved) BCP dipole. $\bm A^{nm} = \langle u_n|\mathrm{i}\nabla_{\bm k}u_m \rangle$ is the Berry connection with $|u_n\rangle$ the periodic part of the $n$th Bloch state, $\epsilon_n$ is the energy of the $n$th Bloch state, $\bm{v}$ is the band velocity, $f(\epsilon_n; \mu)$ is the Fermi-Dirac distribution for energy $\epsilon_n$ at the chemical potential $\mu$, and $d$ is the dimension of the system.

We first analyze the symmetry constraint on $\INH$. Taking $\bm{N}\parallel\hat{x}$ as an example, the magnetic configuration belongs to the $2'/m$ magnetic space group. The allowed components are $\INH^{xzx}=-\INH^{zxx}, \INH^{yzy}=-\INH^{zyy}, \INH^{xyx}=-\INH^{yxx}$, and $\INH^{yzz}=-\INH^{zyz}$, while the rest vanishes \footnotemark[\value{footnote}]. Given that the Hall bar for transport measurements of 2D materials is usually set up within the plane, we focus on the in-plane component $\INH^{yxx}$ ($\INH^{xyy}$) with $\bm{N}$ along the $x$ $(y)$ direction for describing the INH effect in 2D MnS. Figures~\ref{fig2}(a) and~\ref{fig2}(b) show the calculated $\INH$ as a function of the chemical potential $\mu$. For a down-shift of $\mu$ upon hole doping, $\INH^{yxx}$ and $\INH^{xyy}$ exhibit significant peaks with opposite signs at $\mu=-8$ and $-19$ meV near the band edge, and at $-795$, and $-821$ meV which are close to the NDPs along the $\Gamma$-$M$ line. This signifies that the dominant contributions of $\INH$ are from these small-gap regions. Remarkably, when $\bm{N}$ is along the $x$ ($y$) direction, the peaks of $\INH^{yxx}$ ($\INH^{xyy}$) are on the order of nm$\cdot$mA/V$^2$.

It is worth noting that despite the atomically ultrathin 2D nature, the significant value of $\INH$ in MnS is two orders of magnitude larger than the values reported in antiferromagnetic CuMnAs \cite{godinho_electrically_2018, PhysRevLett.127.277201} and Mn$_2$Au \cite{PhysRevLett.127.277202}  ($\sim 10^{-2}$ nm$\cdot$mA/V$^2$), which are prototype materials of antiferromagnetic memory devices. The peak value of $\INH$ in MnS is even comparable to the large $\BCD$ in $\mathcal{T}$-invariant few-layer WTe$_2$ \cite{Kang2019_fewWTe2,Wanghua2019fewWTe2}. In practice, the carrier doping for 2D materials ($\sim 10^{13}$ cm$^{-2}$) can be conveniently controlled by electric gating \cite{ma2019observation, science.1144657, PhysRevLett.105.176602, PhysRevLett.105.256805,nanolett.5b04441}, electron-beam irradiation \cite{Shi2020ElectronBeam, aelm.202100449} or remote modulation \cite{Lee2021RemoteModulation, Zhao2017doping}. We, therefore, expect that it is experimentally feasible to measure our predicted INH effect in the 2D antiferromagnet MnS.

Next, we show that $\INH$ depends sensitively on the direction of $\bm{N}$.
As shown in Figs.~\ref{fig2}(c) and~\ref{fig2}(d), both $\INH^{xyy}$ and $\INH^{yxx}$ exhibit a $2\pi$ periodicity when $\bm{N}$ rotates in the   plane, which satisfies the $\mathcal{T}$-odd constraint that $\INH(\theta)=-\INH(\theta+\pi)$. The angular dependence of $\INH$ is approximately described by cosine or sine trigonometric functions. Therefore, the reorientation of $\bm{N}$ can be fully detected by measuring the INH effect, which is a unique merit over the conventional AMR-based approach that cannot distinguish a 180$^\circ$ reversal.
In particular, the sign of $\INH^{yxx}$ ($\INH^{xyy}$) can be used to distinguish $\bm{N}$ %the N\'eel vector
reversal in $\pm x$ direction ($\theta=0$ or $\pi$) [in $\pm y$ direction ($\theta=\pi/2$ or $3\pi/2$)]. Therefore, the INH effect can serve as a powerful tool for detecting $\bm{N}$.

\begin{figure}
\includegraphics[width =1\columnwidth]{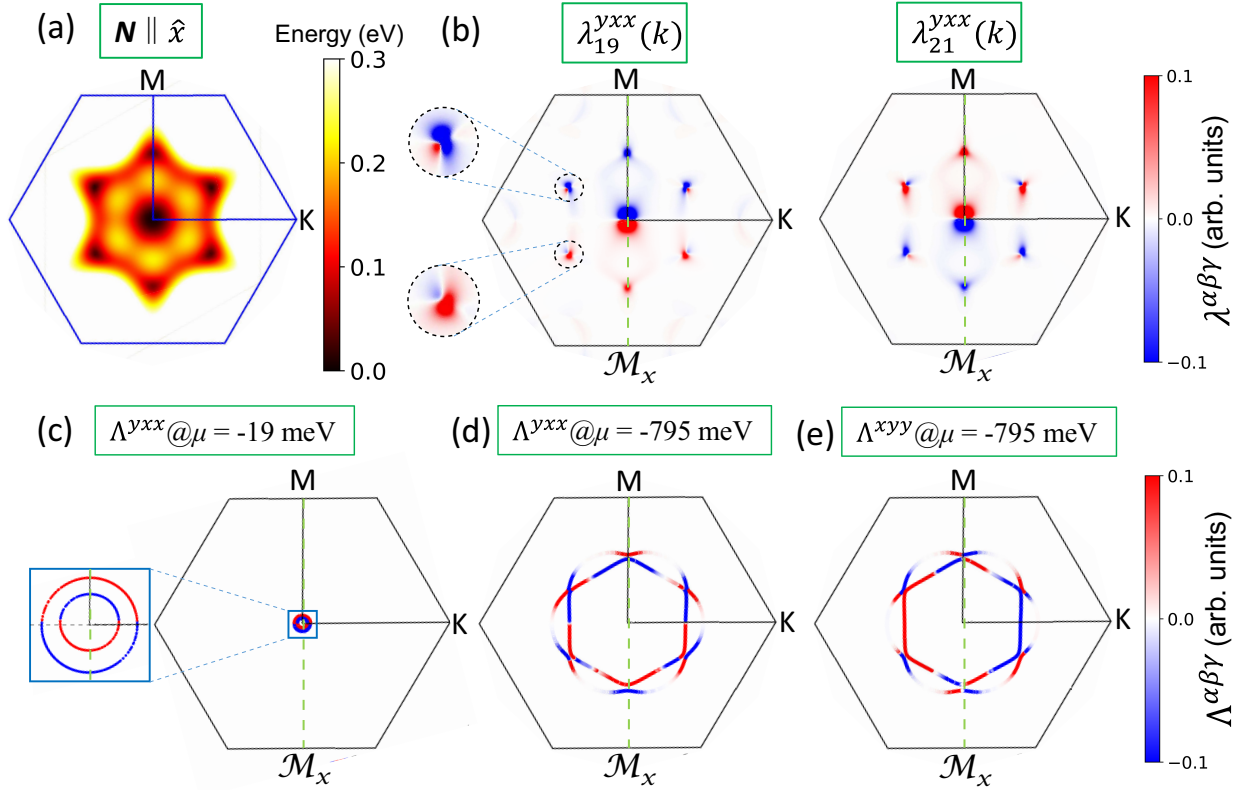}%
\caption{\label{fig4_lambda} (a) Energy difference between top two valence bands in the Brillouin zone. (b) The band-resolved BCP dipole $\lambda^{yxx}_n$ for the top two valence bands ($n=19$ and $21$) for $\bm{N}\parallel\hat{x}$. The distribution around two NDPs are zoomed in. (c)-(e) $k$-resolved distribution of (c) $\Lambda^{yxx}$ at $\mu=-19$ meV, (d) $\Lambda^{yxx}$ at $-795$ meV, and (e) $\Lambda^{xyy}$ at $-795$ meV for $\bm{N}\parallel\hat{x}$. The insert in (c) show the zoom-in plot of $\Lambda^{yxx}$ around $\Gamma$ at $-19$ meV.{The vertical green dashed lines along with the symbol $\mathcal{M}_x$ in (b)-(e) indicate the mirror symmetry perpendicular to the crystalline x axis.}}
\end{figure}

To gain underlying insight into the behavior of $\INH^{yxx}$, we analyze the band-resolved BCP dipole $\lambda^{yxx}_n(k)$, which exhibits the contribution to $\INH^{yxx}$ from each band. {Similar to other band geometric quantities such as the Berry curvature, $\lambda^{yxx}_n(k)$ encodes the interband coherence.} Figure~\ref{fig4_lambda}(a) shows the energy difference between the top two valence bands, where one NDP at $\Gamma$ and six along $\Gamma$-$M$ lines can be observed. In addition, the small-gap region forms a snowflakelike shape centered at $\Gamma$. These NDPs give rise to small denominators for the BCP in Eq.~(\ref{bcp}) and hence a large contribution to $\lambda^{yxx}_n$ for the top two valance bands, as shown in Fig.~\ref{fig4_lambda}(b).

Due to the derivative of the Fermi-Dirac function $\partial f/\partial \epsilon\approx \delta(\epsilon-\mu)$ in Eq.~(\ref{inhexp}), $\INH$ is a Fermi surface property. Therefore, only NDPs close to $\mu$ make significant contributions to $\INH$. To examine the $\bm{k}$-resolved contribution for different $\INH$ peaks, we plot the distribution of the BCP dipole $\Lambda^{yxx}(\bm{k})$ for different peaks of $\INH^{yxx}$, as shown in Figs.~\ref{fig4_lambda}(c) and~\ref{fig4_lambda}(d). The calculated $\Lambda^{yxx}$ at $\mu= -19$ meV mainly distributes around $\Gamma$, while the dominant contribution to $\Lambda^{yxx}$ at $\mu=-795$ meV comes from the small-gap region including the rest NDPs. This indicates that a large $\INH$ can be achieved by tuning $\mu$ towards such regions.

As shown in Fig.~\ref{fig4_lambda}(c)-(e), despite complicated sign changes of $\Lambda^{yxx}$, it is actually an even function with respect to $\Gamma$-$M$ due to the additional constraint from crystalline symmetries $\mathcal{M}_x$ for $\bm{N}\parallel\hat{x}$. On the contrary, $\Lambda^{xyy}$ is dictated to be an odd function with respect to $\mathcal{M}_{x}$ [see Fig.~\ref{fig4_lambda}(e)], which leads to the vanishment of $\INH^{xyy}$ at $\theta=0$ (and $\pi$) [see Fig.~\ref{fig2}(d)]. Similarly, %when $\bm{N}$ is along the $y$ direction,
for $\bm{N}\parallel\hat{y}$, $\Lambda^{xyy}$ becomes an even function with respect to $\Gamma$-$K$, but the preserved symmetry $\mathcal{C}_{2x}$ demands $\Lambda^{yxx}$ to be an odd function (see Fig. S9 \footnotemark[\value{footnote}]). Although the distribution of $\Lambda^{\alpha\beta\gamma}$ seems unaltered when $\bm{N}$ rotates from $\hat{x}$ to $\hat{y}$, our detailed analysis indicates that it delicately changes to satisfy different symmetry constraints, which results in distinct $\INH$ after integrating over the whole Brillouin zone.

\textit{Effective $k\cdot p$ model.}---To better understand the N\'eel vector orientation dependence of $\INH$ in MnS, we construct an effective $k\cdot p$ model to describe the top two valence bands around $\Gamma$. To do so, we first establish an effective model with $D_{3d}$ symmetry and then consider the antiferromagnetism by introducing opposite Zeeman exchange fields for two sublattices. We can start from the antibonding and bonding states of S-$p$ orbitals on two sublattices, $|\eta=\pm,p_\alpha,s\rangle=\frac{1}{\sqrt{2}}(|S_1, p_\alpha,s\rangle \pm |S_2, p_\alpha,s\rangle)$, where $\alpha$ indicates $p_x\pm ip_y$ orbitals and $s=\uparrow\downarrow$ for spin. We label the sub-lattice, orbital, and spin degree of freedom with Pauli matrices $\lambda,\tau$, and $\sigma$, respectively.
The symmetry operations of $D_{3d}$ group are represented as: $\mathcal{C}_{3 z}=\lambda_{0} \otimes \exp \left(-i 2\pi \tau_{z} / 3 \right) \otimes \exp \left(-i \pi \sigma_{z} / 3 \right)$, mirror symmetry $\mathcal{M}_{y}=\lambda_{0} \otimes -\tau_{x} \otimes -i \sigma_{x} $, inversion symmetry $\mathcal{P}=\lambda_{z} \otimes -\tau_{0} \otimes \sigma_{0} $, time reversal symmetry $\mathcal{T}=\lambda_{0} \otimes -\tau_{x} \otimes -i \sigma_{y} K $, where $K$ is complex conjugate operator. In this representation, the full eight-band Hamiltonian reads,
\begin{eqnarray}
\label{C3v 8 band Hamiltonian}
H &=& \begin{pmatrix}
  H_{+} & T_x-iT_y \\
  T_x+iT_y & H_{-}
\end{pmatrix}+J\lambda_x\tau_0(\bm{n}\cdot\bm{\sigma}),
\end{eqnarray}
where $H_\pm$ are antibonding/bonding subspace Hamiltonian, $T_y$ couples two sublattices, and $T_x$ is the coupling of orbitals within one sublattice (see Supplemental Material \footnotemark[\value{footnote}]). The last term represents the Zeeman exchange field where $J$ is the coupling strength and $\bm{n}=(n_x,n_y,n_z)$ represents the N\'eel vector orientation. To describe the top two valence bands, we then downfold the Hamiltonian to the anti-bonding subspace based on the L\"{o}wding perturbation method \cite{winkler2003soc, huang2013existence}, which yields
\begin{eqnarray}
\label{heff}
H_{\mathrm{eff}}&=&H_{D_{3d}}+H_{\bm{n},\parallel} + H_{\bm{n},\perp}, \\
H_{D_{3d}}&=&C_0+C_{1}k^2 +(C_{2}+C_{3}k^2) \tau_{z} \sigma_{z} \nonumber\\
   & &+ C_{4} k_{-}^2\tau_{x} \sigma_{0} + C_{5}k_{+}^2\tau_{z} \sigma_{y},  \nonumber\\
H_{\bm{n},\parallel} &=& A_{1} (k_y n_x - k_x n_y) + A_{2} (k_y n_x - k_x n_y)\tau_z\sigma_z  \nonumber\\
& &+ A_{3} \left[(k_y n_x + k_x n_y) \tau_x + (k_y n_y - k_x n_x) \tau_y\right]\sigma_0, \nonumber\\
H_{\bm{n},\perp} &=& A_{4} (k_y \tau_z\sigma_x -k_x \tau_z\sigma_y)n_z + A_{5} (k_y \tau_y - k_x \tau_x)\sigma_0 n_z, \nonumber
\end{eqnarray}
where $k_\pm=k_x\pm ik_y, k^2=k_x^2+k_y^2$. $C_i$ are material-dependent parameters, among which $C_2$ and $C_3$ represent the SOC induced band splitting. $A_i$ are related to antiferromagnetism.

As a simple illustration, let us consider the case of $\bm{N}\parallel \hat{x}$ [i.e., $\bm{n}=(1,0,0)$]. Keeping up to the lowest order of $k$, we arrive at
\begin{equation}
\label{nx}
\begin{aligned}
H_{\mathrm{eff}}^{(100)} (k)=& C_0 + A_{1} k_y + (C_2 + A_{2} k_y)\tau_z\sigma_z \\
 &+ A_{3} (k_y\tau_x -k_x\tau_y ).
\end{aligned}
\end{equation}
This is a tilted massive Dirac model where $C_2$ is the SOC-induced mass controlling the gap and $A_1$ represents the AFM-induced tile of the Dirac cone. The energy spectrum is given by $E= C_0 + A_{1} k_y \pm \sqrt{(C_2+A_2 k_y)^2 + A_{3}^2  (k_x^2+k_y^2)}$. In this model, the nonvanishing component is $\INH^{yxx} = -\INH^{xyx}$ which exhibits two peaks with opposite signs when $\mu$ approaches the small-gap region, which is consistent with our first-principles calculations in Fig.~\ref{fig2}(a). Because the quadratic terms of $k$, which would bend over the upper valence band significantly, are neglected in the above model (\ref{nx}), the quantitative discrepancy would be eliminated once these terms are included (see Fig. S1-S2 \footnotemark[\value{footnote}]).

More interestingly, when $\bm{N}$ rotates in the \textit{x}-\textit{y} plane with a polar angle $\theta$ with respect to the $x$-axis, it is convenient to obtain the effective model by a coordinate transformation. Consequently, the $\INH$ transform as
\begin{eqnarray}
\label{theta_dependence1}
\INH^{yxx}(\theta)&=& \cos(\theta)\INH^{yxx}(0),\\
\INH^{xyy}(\theta)&=&-\sin(\theta)\INH^{yxx}(0),
\label{theta_dependence2}
\end{eqnarray}
which is consistent with the $\theta$-dependent behavior of $\INH$ in Fig.~\ref{fig2}(c). Similarly, for $\INH$ at $\mu\approx -800$ meV, we construct an effective model with 3 pairs of overtilted massive Dirac cones \cite{huanghqPtSe2,PtTe2} that are related by symmetry, and the angular dependence of $\INH$ is attributed to the $\bm{N}$-dependent tilts that are not canceled within Dirac cone pairs \footnotemark[\value{footnote}]. Thus, the effective $k\cdot p$ model qualitatively explains the large $\INH$ at small-gap regions and its special angular dependence in MnS. These features of the INH effect should be generally expected for 2D $\mathcal{PT}$-symmetric antiferromagnets with similar band structures.

\begin{figure}
\includegraphics[width =1\columnwidth]{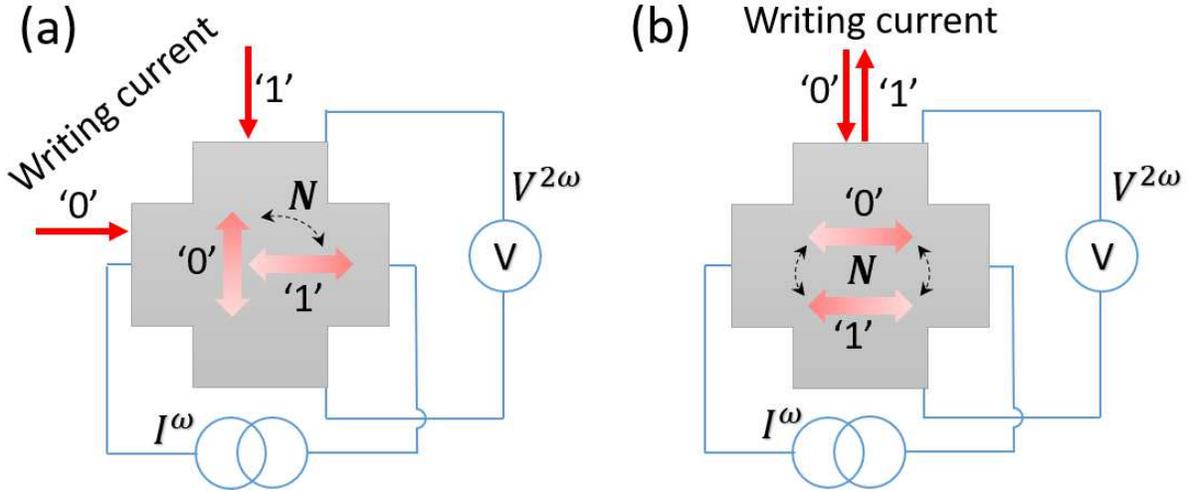}%
\caption{\label{fig5_device}
(a) The 90$^\circ$ $\bm{N}$ switching is controlled by two orthogonal writing currents.
(b) The 180$^\circ$ $\bm{N}$ reversal is controlled by flipping the polarity of the writing current. The write current (red arrows) and the corresponding preferred N\'eel vector orientations (red double-arrows) are labeled `0' and `1'.
The readout is performed by injecting a probing current $I^\omega$ and measuring the nonlinear transverse voltage $V^{2\omega}$. }
\end{figure}

\textit{Discussion and summary.}---In usual Hall measurements with a planar geometry of the setup, the applied electric field may be along a general direction instead of aligning with the crystal axes. When one applies an in-plane electric field $\bm{E}=E(\cos\phi, \sin\phi, 0)$ where $\phi$ is the polar angle with respect to the $x$-axis, the measured in-plane INH current (along the perpendicular direction) is
\begin{equation}
J_\mathrm{INH}=\INH^{\mathrm{in-plane}}(\theta,\phi)E^2,
\end{equation}
where the angle-dependent INH conductivity is
\begin{equation}
\begin{aligned}
\INH^{\mathrm{in-plane}}(\theta,\phi)
=& \cos(\theta-\phi)\INH^{yxx}(0).
\end{aligned}
\end{equation}
The INH conductivity is maximized (minimized) when $\bm{E}$ and $\bm{N}$ are parallel (anti-parallel), but vanishes when they are perpendicular.

The above functionality motivates us to propose a 2D antiferromagnetic memory device based on the standard Hall bar setup. As shown in Fig.~\ref{fig5_device},
reversible orthogonal switching or 180$^\circ$ reversal of $\bm{N}$, which represents two memory states, can be controlled by applying the writing current along two orthogonal directions \cite{wadley_electrical_2016, olejnik_antiferromagnetic_2017, Bodnar_writingAndReadingMn2Au, PhysRevApplied.9.054028,PhysRevApplied.9.064040} or flipping
its polarity \cite{godinho_electrically_2018,wadley-CurrentPolaritydependentManipulation-2018}.
In both schemes, the INH detection of $\bm{N}$ can be performed by injecting a probing current $I^\omega$ with frequency $\omega$ and measuring the transverse voltage with double frequency $V^{2\omega}$, which has been implemented in previous nonlinear Hall measurements \cite{ma2019observation,Kang2019_fewWTe2, PhysRevLett.121.266601, nsr2022_TBWSe2, He2021_TBWTe2,PhysRevLett.129.186801, He2022_grapheneMoire, PhysRevB.106.L041111}. However, the two states in Fig.~\ref{fig5_device}(a) [\ref{fig5_device}(b)] are represented by zero and finite signals (two opposite signals).

In summary, we have predicted the INH detection of $\bm{N}$ %the N\'eel vector
in 2D antiferromagnets Mn$X$, which provides a promising material platform and efficient electric readout approach for 2D antiferromagnetic spintronics. Combined with the high-speed write-in scheme using picosecond current pulses, it is possible to achieve ultra-fast and multi-level memory device applications based on 2D antiferromagnets. For example, a six-level triaxial memory with $\bm{N}$ parallel or anti-parallel to three equivalent axes of MnS can be constructed since it is now able to distinguish these states and their reversed images via the INH effect. In addition, the vast number of 2D antiferromagnetic semiconductors, such as MnPS$_3$ \cite{PhysRevB.94.184428, PhysRevLett.124.027601, PhysRevLett.117.267203, Ni2021imaging}, TaCoTe$_2$ \cite{PhysRevB.100.205102}, Fe$_2$TeO$_6$ and SrFe$_2$S$_2$O \cite{PhysRevLett.129.187602}, bilayer Fe$_3$GeTe$_2$ \cite{PhysRevB.104.104427} and CrCl$_3$ \cite{nanolett.9b01317}, hold great promise for future research.

\begin{acknowledgments}
This work was supported by the National Key R\&D Program of China (Grant No. 2021YFA1401600), the National Natural Science Foundation of China (Grant No. 12074006), and the start-up fund from Peking University. H.Z. and W.D. acknowledge support from the Basic Science Center Project of NSFC (Grant No. 51788104), the Ministry of Science and Technology of China, and the Beijing Advanced Innovation Center for Future Chip (ICFC). The computational resources were supported by the high-performance computing platform of Peking University and the National Supercomputer Center in Guangzhou (NSCC-GZ).

J.W. and H.Z. contributed equally to this work.
\end{acknowledgments}

\bibliography{main}

\begin{thebibliography}{72}%
\makeatletter
\providecommand \@ifxundefined [1]{%
 \@ifx{#1\undefined}
}%
\providecommand \@ifnum [1]{%
 \ifnum #1\expandafter \@firstoftwo
 \else \expandafter \@secondoftwo
 \fi
}%
\providecommand \@ifx [1]{%
 \ifx #1\expandafter \@firstoftwo
 \else \expandafter \@secondoftwo
 \fi
}%
\providecommand \natexlab [1]{#1}%
\providecommand \enquote  [1]{``#1''}%
\providecommand \bibnamefont  [1]{#1}%
\providecommand \bibfnamefont [1]{#1}%
\providecommand \citenamefont [1]{#1}%
\providecommand \href@noop [0]{\@secondoftwo}%
\providecommand \href [0]{\begingroup \@sanitize@url \@href}%
\providecommand \@href[1]{\@@startlink{#1}\@@href}%
\providecommand \@@href[1]{\endgroup#1\@@endlink}%
\providecommand \@sanitize@url [0]{\catcode `\\12\catcode `\$12\catcode
  `\&12\catcode `\#12\catcode `\^12\catcode `\_12\catcode `\%12\relax}%
\providecommand \@@startlink[1]{}%
\providecommand \@@endlink[0]{}%
\providecommand \url  [0]{\begingroup\@sanitize@url \@url }%
\providecommand \@url [1]{\endgroup\@href {#1}{\urlprefix }}%
\providecommand \urlprefix  [0]{URL }%
\providecommand \Eprint [0]{\href }%
\providecommand \doibase [0]{http://dx.doi.org/}%
\providecommand \selectlanguage [0]{\@gobble}%
\providecommand \bibinfo  [0]{\@secondoftwo}%
\providecommand \bibfield  [0]{\@secondoftwo}%
\providecommand \translation [1]{[#1]}%
\providecommand \BibitemOpen [0]{}%
\providecommand \bibitemStop [0]{}%
\providecommand \bibitemNoStop [0]{.\EOS\space}%
\providecommand \EOS [0]{\spacefactor3000\relax}%
\providecommand \BibitemShut  [1]{\csname bibitem#1\endcsname}%
\let\auto@bib@innerbib\@empty
%</preamble>
\bibitem [{\citenamefont {Avsar}\ \emph {et~al.}(2020)\citenamefont {Avsar},
  \citenamefont {Ochoa}, \citenamefont {Guinea}, \citenamefont {\"Ozyilmaz},
  \citenamefont {van Wees},\ and\ \citenamefont
  {Vera-Marun}}]{RevModPhys.92.021003}%
  \BibitemOpen
  \bibfield  {author} {\bibinfo {author} {\bibfnamefont {A.}~\bibnamefont
  {Avsar}}, \bibinfo {author} {\bibfnamefont {H.}~\bibnamefont {Ochoa}},
  \bibinfo {author} {\bibfnamefont {F.}~\bibnamefont {Guinea}}, \bibinfo
  {author} {\bibfnamefont {B.}~\bibnamefont {\"Ozyilmaz}}, \bibinfo {author}
  {\bibfnamefont {B.~J.}\ \bibnamefont {van Wees}}, \ and\ \bibinfo {author}
  {\bibfnamefont {I.~J.}\ \bibnamefont {Vera-Marun}},\ }\bibfield  {title}
  {\enquote {\bibinfo {title} {Colloquium: Spintronics in graphene and other
  two-dimensional materials},}\ }\href {\doibase 10.1103/RevModPhys.92.021003}
  {\bibfield  {journal} {\bibinfo  {journal} {Rev. Mod. Phys.}\ }\textbf
  {\bibinfo {volume} {92}},\ \bibinfo {pages} {021003} (\bibinfo {year}
  {2020})}\BibitemShut {NoStop}%
\bibitem [{\citenamefont {Zhang}\ \emph {et~al.}(2019)\citenamefont {Zhang},
  \citenamefont {Wong}, \citenamefont {Zhu},\ and\ \citenamefont
  {Wee}}]{inf2.12048}%
  \BibitemOpen
  \bibfield  {author} {\bibinfo {author} {\bibfnamefont {Wen}\ \bibnamefont
  {Zhang}}, \bibinfo {author} {\bibfnamefont {Ping Kwan~Johnny}\ \bibnamefont
  {Wong}}, \bibinfo {author} {\bibfnamefont {Rui}\ \bibnamefont {Zhu}}, \ and\
  \bibinfo {author} {\bibfnamefont {Andrew T.~S.}\ \bibnamefont {Wee}},\
  }\bibfield  {title} {\enquote {\bibinfo {title} {Van der waals magnets:
  Wonder building blocks for two-dimensional spintronics?}}\ }\href {\doibase
  https://doi.org/10.1002/inf2.12048} {\bibfield  {journal} {\bibinfo
  {journal} {InfoMat}\ }\textbf {\bibinfo {volume} {1}},\ \bibinfo {pages}
  {479--495} (\bibinfo {year} {2019})}\BibitemShut {NoStop}%
\bibitem [{\citenamefont {Li}\ and\ \citenamefont {Wu}(2016)}]{wcms.1259}%
  \BibitemOpen
  \bibfield  {author} {\bibinfo {author} {\bibfnamefont {Xiuling}\ \bibnamefont
  {Li}}\ and\ \bibinfo {author} {\bibfnamefont {Xiaojun}\ \bibnamefont {Wu}},\
  }\bibfield  {title} {\enquote {\bibinfo {title} {Two-dimensional monolayer
  designs for spintronics applications},}\ }\href {\doibase
  https://doi.org/10.1002/wcms.1259} {\bibfield  {journal} {\bibinfo  {journal}
  {WIRES: Comp. Mol. Sci.}\ }\textbf {\bibinfo {volume} {6}},\ \bibinfo {pages}
  {441--455} (\bibinfo {year} {2016})}\BibitemShut {NoStop}%
\bibitem [{\citenamefont {Lin}\ \emph {et~al.}(2019)\citenamefont {Lin},
  \citenamefont {Yang}, \citenamefont {Wang},\ and\ \citenamefont
  {Zhao}}]{linXiaoyang2019}%
  \BibitemOpen
  \bibfield  {author} {\bibinfo {author} {\bibfnamefont {Xiaoyang}\
  \bibnamefont {Lin}}, \bibinfo {author} {\bibfnamefont {Wei}\ \bibnamefont
  {Yang}}, \bibinfo {author} {\bibfnamefont {Kang~L.}\ \bibnamefont {Wang}}, \
  and\ \bibinfo {author} {\bibfnamefont {Weisheng}\ \bibnamefont {Zhao}},\
  }\bibfield  {title} {\enquote {\bibinfo {title} {Two-dimensional spintronics
  for low-power electronics},}\ }\href {\doibase 10.1038/s41928-019-0273-7}
  {\bibfield  {journal} {\bibinfo  {journal} {Nature Electronics}\ }\textbf
  {\bibinfo {volume} {2}},\ \bibinfo {pages} {274–283} (\bibinfo {year}
  {2019})}\BibitemShut {NoStop}%
\bibitem [{\citenamefont {Ahn}(2020)}]{AhnEthan2020}%
  \BibitemOpen
  \bibfield  {author} {\bibinfo {author} {\bibfnamefont {Ethan~C.}\
  \bibnamefont {Ahn}},\ }\bibfield  {title} {\enquote {\bibinfo {title} {2d
  materials for spintronic devices},}\ }\href {\doibase
  10.1038/s41699-020-0152-0} {\bibfield  {journal} {\bibinfo  {journal} {npj 2D
  Materials and Applications}\ }\textbf {\bibinfo {volume} {4}},\ \bibinfo
  {pages} {17} (\bibinfo {year} {2020})}\BibitemShut {NoStop}%
\bibitem [{\citenamefont {Han}(2016)}]{Hanwei2016}%
  \BibitemOpen
  \bibfield  {author} {\bibinfo {author} {\bibfnamefont {Wei}\ \bibnamefont
  {Han}},\ }\bibfield  {title} {\enquote {\bibinfo {title} {Perspectives for
  spintronics in 2d materials},}\ }\href {\doibase 10.1063/1.4941712}
  {\bibfield  {journal} {\bibinfo  {journal} {APL Materials}\ }\textbf
  {\bibinfo {volume} {4}},\ \bibinfo {pages} {032401} (\bibinfo {year}
  {2016})}\BibitemShut {NoStop}%
\bibitem [{\citenamefont {Xiao}\ \emph {et~al.}(2012)\citenamefont {Xiao},
  \citenamefont {Liu}, \citenamefont {Feng}, \citenamefont {Xu},\ and\
  \citenamefont {Yao}}]{PhysRevLett.108.196802}%
  \BibitemOpen
  \bibfield  {author} {\bibinfo {author} {\bibfnamefont {Di}~\bibnamefont
  {Xiao}}, \bibinfo {author} {\bibfnamefont {Gui-Bin}\ \bibnamefont {Liu}},
  \bibinfo {author} {\bibfnamefont {Wanxiang}\ \bibnamefont {Feng}}, \bibinfo
  {author} {\bibfnamefont {Xiaodong}\ \bibnamefont {Xu}}, \ and\ \bibinfo
  {author} {\bibfnamefont {Wang}\ \bibnamefont {Yao}},\ }\bibfield  {title}
  {\enquote {\bibinfo {title} {Coupled spin and valley physics in monolayers of
  ${\mathrm{mos}}_{2}$ and other group-vi dichalcogenides},}\ }\href {\doibase
  10.1103/PhysRevLett.108.196802} {\bibfield  {journal} {\bibinfo  {journal}
  {Phys. Rev. Lett.}\ }\textbf {\bibinfo {volume} {108}},\ \bibinfo {pages}
  {196802} (\bibinfo {year} {2012})}\BibitemShut {NoStop}%
\bibitem [{\citenamefont {Kane}\ and\ \citenamefont
  {Mele}(2005)}]{PhysRevLett.95.226801}%
  \BibitemOpen
  \bibfield  {author} {\bibinfo {author} {\bibfnamefont {C.~L.}\ \bibnamefont
  {Kane}}\ and\ \bibinfo {author} {\bibfnamefont {E.~J.}\ \bibnamefont
  {Mele}},\ }\bibfield  {title} {\enquote {\bibinfo {title} {Quantum spin hall
  effect in graphene},}\ }\href {\doibase 10.1103/PhysRevLett.95.226801}
  {\bibfield  {journal} {\bibinfo  {journal} {Phys. Rev. Lett.}\ }\textbf
  {\bibinfo {volume} {95}},\ \bibinfo {pages} {226801} (\bibinfo {year}
  {2005})}\BibitemShut {NoStop}%
\bibitem [{\citenamefont {Zhang}\ \emph {et~al.}(2021)\citenamefont {Zhang},
  \citenamefont {Xu}, \citenamefont {Luo},\ and\ \citenamefont
  {Zou}}]{ZouXiaolong2021}%
  \BibitemOpen
  \bibfield  {author} {\bibinfo {author} {\bibfnamefont {Shuqing}\ \bibnamefont
  {Zhang}}, \bibinfo {author} {\bibfnamefont {Runzhang}\ \bibnamefont {Xu}},
  \bibinfo {author} {\bibfnamefont {Nannan}\ \bibnamefont {Luo}}, \ and\
  \bibinfo {author} {\bibfnamefont {Xiaolong}\ \bibnamefont {Zou}},\ }\bibfield
   {title} {\enquote {\bibinfo {title} {Two-dimensional magnetic materials:
  structures{,} properties and external controls},}\ }\href {\doibase
  10.1039/D0NR06813F} {\bibfield  {journal} {\bibinfo  {journal} {Nanoscale}\
  }\textbf {\bibinfo {volume} {13}},\ \bibinfo {pages} {1398--1424} (\bibinfo
  {year} {2021})}\BibitemShut {NoStop}%
\bibitem [{\citenamefont {Huang}\ \emph {et~al.}(2017)\citenamefont {Huang},
  \citenamefont {Clark}, \citenamefont {Navarro-Moratalla}, \citenamefont
  {Klein}, \citenamefont {Cheng}, \citenamefont {Seyler}, \citenamefont
  {Zhong}, \citenamefont {Schmidgall}, \citenamefont {McGuire}, \citenamefont
  {Cobden}, \citenamefont {Yao}, \citenamefont {Xiao}, \citenamefont
  {Jarillo-Herrero},\ and\ \citenamefont {Xu}}]{Bevin2017CrI3}%
  \BibitemOpen
  \bibfield  {author} {\bibinfo {author} {\bibfnamefont {Bevin}\ \bibnamefont
  {Huang}}, \bibinfo {author} {\bibfnamefont {Genevieve}\ \bibnamefont
  {Clark}}, \bibinfo {author} {\bibfnamefont {Efrén}\ \bibnamefont
  {Navarro-Moratalla}}, \bibinfo {author} {\bibfnamefont {Dahlia~R.}\
  \bibnamefont {Klein}}, \bibinfo {author} {\bibfnamefont {Ran}\ \bibnamefont
  {Cheng}}, \bibinfo {author} {\bibfnamefont {Kyle~L.}\ \bibnamefont {Seyler}},
  \bibinfo {author} {\bibfnamefont {Ding}\ \bibnamefont {Zhong}}, \bibinfo
  {author} {\bibfnamefont {Emma}\ \bibnamefont {Schmidgall}}, \bibinfo {author}
  {\bibfnamefont {Michael~A.}\ \bibnamefont {McGuire}}, \bibinfo {author}
  {\bibfnamefont {David~H.}\ \bibnamefont {Cobden}}, \bibinfo {author}
  {\bibfnamefont {Wang}\ \bibnamefont {Yao}}, \bibinfo {author} {\bibfnamefont
  {Di}~\bibnamefont {Xiao}}, \bibinfo {author} {\bibfnamefont {Pablo}\
  \bibnamefont {Jarillo-Herrero}}, \ and\ \bibinfo {author} {\bibfnamefont
  {Xiaodong}\ \bibnamefont {Xu}},\ }\bibfield  {title} {\enquote {\bibinfo
  {title} {Layer-dependent ferromagnetism in a van der waals crystal down to
  the monolayer limit},}\ }\href {\doibase 10.1038/nature22391} {\bibfield
  {journal} {\bibinfo  {journal} {Nature}\ }\textbf {\bibinfo {volume} {546}},\
  \bibinfo {pages} {270--273} (\bibinfo {year} {2017})}\BibitemShut {NoStop}%
\bibitem [{\citenamefont {Gong}\ \emph {et~al.}(2017)\citenamefont {Gong},
  \citenamefont {Li}, \citenamefont {Li}, \citenamefont {Ji}, \citenamefont
  {Stern}, \citenamefont {Xia}, \citenamefont {Cao}, \citenamefont {Bao},
  \citenamefont {Wang}, \citenamefont {Wang}, \citenamefont {Qiu},
  \citenamefont {Cava}, \citenamefont {Louie}, \citenamefont {Xia},\ and\
  \citenamefont {Zhang}}]{Gong2017Cr2Ge2Te6}%
  \BibitemOpen
  \bibfield  {author} {\bibinfo {author} {\bibfnamefont {Cheng}\ \bibnamefont
  {Gong}}, \bibinfo {author} {\bibfnamefont {Lin}\ \bibnamefont {Li}}, \bibinfo
  {author} {\bibfnamefont {Zhenglu}\ \bibnamefont {Li}}, \bibinfo {author}
  {\bibfnamefont {Huiwen}\ \bibnamefont {Ji}}, \bibinfo {author} {\bibfnamefont
  {Alex}\ \bibnamefont {Stern}}, \bibinfo {author} {\bibfnamefont {Yang}\
  \bibnamefont {Xia}}, \bibinfo {author} {\bibfnamefont {Ting}\ \bibnamefont
  {Cao}}, \bibinfo {author} {\bibfnamefont {Wei}\ \bibnamefont {Bao}}, \bibinfo
  {author} {\bibfnamefont {Chenzhe}\ \bibnamefont {Wang}}, \bibinfo {author}
  {\bibfnamefont {Yuan}\ \bibnamefont {Wang}}, \bibinfo {author} {\bibfnamefont
  {Z.~Q.}\ \bibnamefont {Qiu}}, \bibinfo {author} {\bibfnamefont {R.~J.}\
  \bibnamefont {Cava}}, \bibinfo {author} {\bibfnamefont {Steven~G.}\
  \bibnamefont {Louie}}, \bibinfo {author} {\bibfnamefont {Jing}\ \bibnamefont
  {Xia}}, \ and\ \bibinfo {author} {\bibfnamefont {Xiang}\ \bibnamefont
  {Zhang}},\ }\bibfield  {title} {\enquote {\bibinfo {title} {Discovery of
  intrinsic ferromagnetism in two-dimensional van der waals crystals},}\ }\href
  {\doibase 10.1038/nature22060} {\bibfield  {journal} {\bibinfo  {journal}
  {Nature}\ }\textbf {\bibinfo {volume} {546}},\ \bibinfo {pages} {265--269}
  (\bibinfo {year} {2017})}\BibitemShut {NoStop}%
\bibitem [{\citenamefont {Baltz}\ \emph {et~al.}(2018)\citenamefont {Baltz},
  \citenamefont {Manchon}, \citenamefont {Tsoi}, \citenamefont {Moriyama},
  \citenamefont {Ono},\ and\ \citenamefont
  {Tserkovnyak}}]{Baltz.2018.Tserkovnyak}%
  \BibitemOpen
  \bibfield  {author} {\bibinfo {author} {\bibfnamefont {V.}~\bibnamefont
  {Baltz}}, \bibinfo {author} {\bibfnamefont {A.}~\bibnamefont {Manchon}},
  \bibinfo {author} {\bibfnamefont {M.}~\bibnamefont {Tsoi}}, \bibinfo {author}
  {\bibfnamefont {T.}~\bibnamefont {Moriyama}}, \bibinfo {author}
  {\bibfnamefont {T.}~\bibnamefont {Ono}}, \ and\ \bibinfo {author}
  {\bibfnamefont {Y.}~\bibnamefont {Tserkovnyak}},\ }\bibfield  {title}
  {\enquote {\bibinfo {title} {{Antiferromagnetic spintronics}},}\ }\href
  {\doibase 10.1103/revmodphys.90.015005} {\bibfield  {journal} {\bibinfo
  {journal} {Rev. Mod. Phys.}\ }\textbf {\bibinfo {volume} {90}},\ \bibinfo
  {pages} {015005} (\bibinfo {year} {2018})}\BibitemShut {NoStop}%
\bibitem [{\citenamefont {Jungwirth}\ \emph {et~al.}(2016)\citenamefont
  {Jungwirth}, \citenamefont {Marti}, \citenamefont {Wadley},\ and\
  \citenamefont {Wunderlich}}]{jungwirth_antiferromagnetic_2016}%
  \BibitemOpen
  \bibfield  {author} {\bibinfo {author} {\bibfnamefont {T.}~\bibnamefont
  {Jungwirth}}, \bibinfo {author} {\bibfnamefont {X.}~\bibnamefont {Marti}},
  \bibinfo {author} {\bibfnamefont {P.}~\bibnamefont {Wadley}}, \ and\ \bibinfo
  {author} {\bibfnamefont {J.}~\bibnamefont {Wunderlich}},\ }\bibfield  {title}
  {\enquote {\bibinfo {title} {Antiferromagnetic spintronics},}\ }\href
  {\doibase 10.1038/nnano.2016.18} {\bibfield  {journal} {\bibinfo  {journal}
  {Nat. Nanotechnol.}\ }\textbf {\bibinfo {volume} {11}},\ \bibinfo {pages}
  {231--241} (\bibinfo {year} {2016})}\BibitemShut {NoStop}%
\bibitem [{\citenamefont {\ifmmode~\check{Z}\else \v{Z}\fi{}elezn\'y}\ \emph
  {et~al.}(2014)\citenamefont {\ifmmode~\check{Z}\else \v{Z}\fi{}elezn\'y},
  \citenamefont {Gao}, \citenamefont {V\'yborn\'y}, \citenamefont {Zemen},
  \citenamefont {Ma\ifmmode~\check{s}\else \v{s}\fi{}ek}, \citenamefont
  {Manchon}, \citenamefont {Wunderlich}, \citenamefont {Sinova},\ and\
  \citenamefont {Jungwirth}}]{PhysRevLett.113.157201}%
  \BibitemOpen
  \bibfield  {author} {\bibinfo {author} {\bibfnamefont {J.}~\bibnamefont
  {\ifmmode~\check{Z}\else \v{Z}\fi{}elezn\'y}}, \bibinfo {author}
  {\bibfnamefont {H.}~\bibnamefont {Gao}}, \bibinfo {author} {\bibfnamefont
  {K.}~\bibnamefont {V\'yborn\'y}}, \bibinfo {author} {\bibfnamefont
  {J.}~\bibnamefont {Zemen}}, \bibinfo {author} {\bibfnamefont
  {J.}~\bibnamefont {Ma\ifmmode~\check{s}\else \v{s}\fi{}ek}}, \bibinfo
  {author} {\bibfnamefont {Aur\'elien}\ \bibnamefont {Manchon}}, \bibinfo
  {author} {\bibfnamefont {J.}~\bibnamefont {Wunderlich}}, \bibinfo {author}
  {\bibfnamefont {Jairo}\ \bibnamefont {Sinova}}, \ and\ \bibinfo {author}
  {\bibfnamefont {T.}~\bibnamefont {Jungwirth}},\ }\bibfield  {title} {\enquote
  {\bibinfo {title} {Relativistic n\'eel-order fields induced by electrical
  current in antiferromagnets},}\ }\href {\doibase
  10.1103/PhysRevLett.113.157201} {\bibfield  {journal} {\bibinfo  {journal}
  {Phys. Rev. Lett.}\ }\textbf {\bibinfo {volume} {113}},\ \bibinfo {pages}
  {157201} (\bibinfo {year} {2014})}\BibitemShut {NoStop}%
\bibitem [{\citenamefont {Manchon}\ \emph {et~al.}(2019)\citenamefont
  {Manchon}, \citenamefont {\ifmmode~\check{Z}\else \v{Z}\fi{}elezn\'y},
  \citenamefont {Miron}, \citenamefont {Jungwirth}, \citenamefont {Sinova},
  \citenamefont {Thiaville}, \citenamefont {Garello},\ and\ \citenamefont
  {Gambardella}}]{RevModPhys.91.035004}%
  \BibitemOpen
  \bibfield  {author} {\bibinfo {author} {\bibfnamefont {A.}~\bibnamefont
  {Manchon}}, \bibinfo {author} {\bibfnamefont {J.}~\bibnamefont
  {\ifmmode~\check{Z}\else \v{Z}\fi{}elezn\'y}}, \bibinfo {author}
  {\bibfnamefont {I.~M.}\ \bibnamefont {Miron}}, \bibinfo {author}
  {\bibfnamefont {T.}~\bibnamefont {Jungwirth}}, \bibinfo {author}
  {\bibfnamefont {J.}~\bibnamefont {Sinova}}, \bibinfo {author} {\bibfnamefont
  {A.}~\bibnamefont {Thiaville}}, \bibinfo {author} {\bibfnamefont
  {K.}~\bibnamefont {Garello}}, \ and\ \bibinfo {author} {\bibfnamefont
  {P.}~\bibnamefont {Gambardella}},\ }\bibfield  {title} {\enquote {\bibinfo
  {title} {Current-induced spin-orbit torques in ferromagnetic and
  antiferromagnetic systems},}\ }\href {\doibase 10.1103/RevModPhys.91.035004}
  {\bibfield  {journal} {\bibinfo  {journal} {Rev. Mod. Phys.}\ }\textbf
  {\bibinfo {volume} {91}},\ \bibinfo {pages} {035004} (\bibinfo {year}
  {2019})}\BibitemShut {NoStop}%
\bibitem [{\citenamefont {Wadley}\ \emph {et~al.}(2016)\citenamefont {Wadley},
  \citenamefont {Howells}, \citenamefont {\v{Z}elezn\'{y}}, \citenamefont
  {Andrews}, \citenamefont {Hills}, \citenamefont {Campion}, \citenamefont
  {Nov\'{a}k}, \citenamefont {Olejn\'{\i}k}, \citenamefont {Maccherozzi},
  \citenamefont {Dhesi}, \citenamefont {Martin}, \citenamefont {Wagner},
  \citenamefont {Wunderlich}, \citenamefont {Freimuth}, \citenamefont
  {Mokrousov}, \citenamefont {Kune\v{s}}, \citenamefont {Chauhan},
  \citenamefont {Grzybowski}, \citenamefont {Rushforth}, \citenamefont
  {Edmonds}, \citenamefont {Gallagher},\ and\ \citenamefont
  {Jungwirth}}]{wadley_electrical_2016}%
  \BibitemOpen
  \bibfield  {author} {\bibinfo {author} {\bibfnamefont {P.}~\bibnamefont
  {Wadley}}, \bibinfo {author} {\bibfnamefont {B.}~\bibnamefont {Howells}},
  \bibinfo {author} {\bibfnamefont {J.}~\bibnamefont {\v{Z}elezn\'{y}}},
  \bibinfo {author} {\bibfnamefont {C.}~\bibnamefont {Andrews}}, \bibinfo
  {author} {\bibfnamefont {V.}~\bibnamefont {Hills}}, \bibinfo {author}
  {\bibfnamefont {R.~P.}\ \bibnamefont {Campion}}, \bibinfo {author}
  {\bibfnamefont {V.}~\bibnamefont {Nov\'{a}k}}, \bibinfo {author}
  {\bibfnamefont {K.}~\bibnamefont {Olejn\'{\i}k}}, \bibinfo {author}
  {\bibfnamefont {F.}~\bibnamefont {Maccherozzi}}, \bibinfo {author}
  {\bibfnamefont {S.~S.}\ \bibnamefont {Dhesi}}, \bibinfo {author}
  {\bibfnamefont {S.~Y.}\ \bibnamefont {Martin}}, \bibinfo {author}
  {\bibfnamefont {T.}~\bibnamefont {Wagner}}, \bibinfo {author} {\bibfnamefont
  {J.}~\bibnamefont {Wunderlich}}, \bibinfo {author} {\bibfnamefont
  {F.}~\bibnamefont {Freimuth}}, \bibinfo {author} {\bibfnamefont
  {Y.}~\bibnamefont {Mokrousov}}, \bibinfo {author} {\bibfnamefont
  {J.}~\bibnamefont {Kune\v{s}}}, \bibinfo {author} {\bibfnamefont {J.~S.}\
  \bibnamefont {Chauhan}}, \bibinfo {author} {\bibfnamefont {M.~J.}\
  \bibnamefont {Grzybowski}}, \bibinfo {author} {\bibfnamefont {A.~W.}\
  \bibnamefont {Rushforth}}, \bibinfo {author} {\bibfnamefont {K.~W.}\
  \bibnamefont {Edmonds}}, \bibinfo {author} {\bibfnamefont {B.~L.}\
  \bibnamefont {Gallagher}}, \ and\ \bibinfo {author} {\bibfnamefont
  {T.}~\bibnamefont {Jungwirth}},\ }\bibfield  {title} {\enquote {\bibinfo
  {title} {Electrical switching of an antiferromagnet},}\ }\href {\doibase
  10.1126/science.aab1031} {\bibfield  {journal} {\bibinfo  {journal}
  {Science}\ }\textbf {\bibinfo {volume} {351}},\ \bibinfo {pages} {587--590}
  (\bibinfo {year} {2016})}\BibitemShut {NoStop}%
\bibitem [{\citenamefont {Olejn\'{\i}k}\ \emph {et~al.}(2017)\citenamefont
  {Olejn\'{\i}k}, \citenamefont {Schuler}, \citenamefont {Marti}, \citenamefont
  {Nov\'{a}k}, \citenamefont {Ka\v{s}par}, \citenamefont {Wadley},
  \citenamefont {Campion}, \citenamefont {Edmonds}, \citenamefont {Gallagher},
  \citenamefont {Garces}, \citenamefont {Baumgartner}, \citenamefont
  {Gambardella},\ and\ \citenamefont
  {Jungwirth}}]{olejnik_antiferromagnetic_2017}%
  \BibitemOpen
  \bibfield  {author} {\bibinfo {author} {\bibfnamefont {K.}~\bibnamefont
  {Olejn\'{\i}k}}, \bibinfo {author} {\bibfnamefont {V.}~\bibnamefont
  {Schuler}}, \bibinfo {author} {\bibfnamefont {X.}~\bibnamefont {Marti}},
  \bibinfo {author} {\bibfnamefont {V.}~\bibnamefont {Nov\'{a}k}}, \bibinfo
  {author} {\bibfnamefont {Z.}~\bibnamefont {Ka\v{s}par}}, \bibinfo {author}
  {\bibfnamefont {P.}~\bibnamefont {Wadley}}, \bibinfo {author} {\bibfnamefont
  {R.~P.}\ \bibnamefont {Campion}}, \bibinfo {author} {\bibfnamefont {K.~W.}\
  \bibnamefont {Edmonds}}, \bibinfo {author} {\bibfnamefont {B.~L.}\
  \bibnamefont {Gallagher}}, \bibinfo {author} {\bibfnamefont {J.}~\bibnamefont
  {Garces}}, \bibinfo {author} {\bibfnamefont {M.}~\bibnamefont {Baumgartner}},
  \bibinfo {author} {\bibfnamefont {P.}~\bibnamefont {Gambardella}}, \ and\
  \bibinfo {author} {\bibfnamefont {T.}~\bibnamefont {Jungwirth}},\ }\bibfield
  {title} {\enquote {\bibinfo {title} {Antiferromagnetic {CuMnAs} multi-level
  memory cell with microelectronic compatibility},}\ }\href {\doibase
  10.1038/ncomms15434} {\bibfield  {journal} {\bibinfo  {journal} {Nat.
  Commun.}\ }\textbf {\bibinfo {volume} {8}},\ \bibinfo {pages} {15434}
  (\bibinfo {year} {2017})}\BibitemShut {NoStop}%
\bibitem [{\citenamefont {Bodnar}\ \emph {et~al.}(2018)\citenamefont {Bodnar},
  \citenamefont {\v{S}mejkal}, \citenamefont {Turek}, \citenamefont
  {Jungwirth}, \citenamefont {Gomonay}, \citenamefont {Sinova}, \citenamefont
  {Sapozhnik}, \citenamefont {Elmers}, \citenamefont {Kl\"{a}ui},\ and\
  \citenamefont {Jourdan}}]{Bodnar_writingAndReadingMn2Au}%
  \BibitemOpen
  \bibfield  {author} {\bibinfo {author} {\bibfnamefont {S.~Yu}\ \bibnamefont
  {Bodnar}}, \bibinfo {author} {\bibfnamefont {L.}~\bibnamefont {\v{S}mejkal}},
  \bibinfo {author} {\bibfnamefont {I.}~\bibnamefont {Turek}}, \bibinfo
  {author} {\bibfnamefont {T.}~\bibnamefont {Jungwirth}}, \bibinfo {author}
  {\bibfnamefont {O.}~\bibnamefont {Gomonay}}, \bibinfo {author} {\bibfnamefont
  {J.}~\bibnamefont {Sinova}}, \bibinfo {author} {\bibfnamefont {A.~A.}\
  \bibnamefont {Sapozhnik}}, \bibinfo {author} {\bibfnamefont {H.~J.}\
  \bibnamefont {Elmers}}, \bibinfo {author} {\bibfnamefont {M.}~\bibnamefont
  {Kl\"{a}ui}}, \ and\ \bibinfo {author} {\bibfnamefont {M.}~\bibnamefont
  {Jourdan}},\ }\bibfield  {title} {\enquote {\bibinfo {title} {Writing and
  reading antiferromagnetic mn$_2$au by n\'eel spin-orbit torques and large
  anisotropic magnetoresistance},}\ }\href {\doibase
  10.1038/s41467-017-02780-x} {\bibfield  {journal} {\bibinfo  {journal} {Nat.
  Commun.}\ }\textbf {\bibinfo {volume} {9}},\ \bibinfo {pages} {348} (\bibinfo
  {year} {2018})}\BibitemShut {NoStop}%
\bibitem [{\citenamefont {Roy}\ \emph {et~al.}(2016)\citenamefont {Roy},
  \citenamefont {Otxoa},\ and\ \citenamefont
  {Wunderlich}}]{PhysRevB.94.014439}%
  \BibitemOpen
  \bibfield  {author} {\bibinfo {author} {\bibfnamefont {P.~E.}\ \bibnamefont
  {Roy}}, \bibinfo {author} {\bibfnamefont {R.~M.}\ \bibnamefont {Otxoa}}, \
  and\ \bibinfo {author} {\bibfnamefont {J.}~\bibnamefont {Wunderlich}},\
  }\bibfield  {title} {\enquote {\bibinfo {title} {Robust picosecond writing of
  a layered antiferromagnet by staggered spin-orbit fields},}\ }\href {\doibase
  10.1103/PhysRevB.94.014439} {\bibfield  {journal} {\bibinfo  {journal} {Phys.
  Rev. B}\ }\textbf {\bibinfo {volume} {94}},\ \bibinfo {pages} {014439}
  (\bibinfo {year} {2016})}\BibitemShut {NoStop}%
\bibitem [{\citenamefont {Olejn\'{\i}k}\ \emph {et~al.}(2018)\citenamefont
  {Olejn\'{\i}k}, \citenamefont {Seifert}, \citenamefont {Ka\v{s}par},
  \citenamefont {Nov\'{a}k}, \citenamefont {Wadley}, \citenamefont {Campion},
  \citenamefont {Baumgartner}, \citenamefont {Gambardella}, \citenamefont
  {N\v{e}mec}, \citenamefont {Wunderlich}, \citenamefont {Sinova},
  \citenamefont {P.}, \citenamefont {M\"{u}ller}, \citenamefont {Kampfrath},\
  and\ \citenamefont {Jungwirth}}]{Olejnik2017ThzWriting}%
  \BibitemOpen
  \bibfield  {author} {\bibinfo {author} {\bibfnamefont {K.}~\bibnamefont
  {Olejn\'{\i}k}}, \bibinfo {author} {\bibfnamefont {T.}~\bibnamefont
  {Seifert}}, \bibinfo {author} {\bibfnamefont {Z.}~\bibnamefont {Ka\v{s}par}},
  \bibinfo {author} {\bibfnamefont {V.}~\bibnamefont {Nov\'{a}k}}, \bibinfo
  {author} {\bibfnamefont {P.}~\bibnamefont {Wadley}}, \bibinfo {author}
  {\bibfnamefont {R.~P.}\ \bibnamefont {Campion}}, \bibinfo {author}
  {\bibfnamefont {M.}~\bibnamefont {Baumgartner}}, \bibinfo {author}
  {\bibfnamefont {P.}~\bibnamefont {Gambardella}}, \bibinfo {author}
  {\bibfnamefont {P.}~\bibnamefont {N\v{e}mec}}, \bibinfo {author}
  {\bibfnamefont {J.}~\bibnamefont {Wunderlich}}, \bibinfo {author}
  {\bibfnamefont {J.}~\bibnamefont {Sinova}}, \bibinfo {author} {\bibfnamefont
  {Ku\v{z}el}\ \bibnamefont {P.}}, \bibinfo {author} {\bibfnamefont
  {M.}~\bibnamefont {M\"{u}ller}}, \bibinfo {author} {\bibfnamefont
  {T.}~\bibnamefont {Kampfrath}}, \ and\ \bibinfo {author} {\bibfnamefont
  {T}~\bibnamefont {Jungwirth}},\ }\bibfield  {title} {\enquote {\bibinfo
  {title} {Terahertz electrical writing speed in an antiferromagnetic
  memory},}\ }\href {\doibase 10.1126/sciadv.aar3566} {\bibfield  {journal}
  {\bibinfo  {journal} {Sci. Adv.}\ }\textbf {\bibinfo {volume} {4}},\ \bibinfo
  {pages} {eaar3566} (\bibinfo {year} {2018})}\BibitemShut {NoStop}%
\bibitem [{\citenamefont {Wadley}\ \emph {et~al.}(2018)\citenamefont {Wadley},
  \citenamefont {Reimers}, \citenamefont {Grzybowski}, \citenamefont {Andrews},
  \citenamefont {Wang}, \citenamefont {Chauhan}, \citenamefont {Gallagher},
  \citenamefont {Campion}, \citenamefont {Edmonds}, \citenamefont {Dhesi},
  \citenamefont {Maccherozzi}, \citenamefont {Novak}, \citenamefont
  {Wunderlich},\ and\ \citenamefont
  {Jungwirth}}]{wadley-CurrentPolaritydependentManipulation-2018}%
  \BibitemOpen
  \bibfield  {author} {\bibinfo {author} {\bibfnamefont {Peter}\ \bibnamefont
  {Wadley}}, \bibinfo {author} {\bibfnamefont {Sonka}\ \bibnamefont {Reimers}},
  \bibinfo {author} {\bibfnamefont {Michal~J.}\ \bibnamefont {Grzybowski}},
  \bibinfo {author} {\bibfnamefont {Carl}\ \bibnamefont {Andrews}}, \bibinfo
  {author} {\bibfnamefont {Mu}~\bibnamefont {Wang}}, \bibinfo {author}
  {\bibfnamefont {Jasbinder~S.}\ \bibnamefont {Chauhan}}, \bibinfo {author}
  {\bibfnamefont {Bryan~L.}\ \bibnamefont {Gallagher}}, \bibinfo {author}
  {\bibfnamefont {Richard~P.}\ \bibnamefont {Campion}}, \bibinfo {author}
  {\bibfnamefont {Kevin~W.}\ \bibnamefont {Edmonds}}, \bibinfo {author}
  {\bibfnamefont {Sarnjeet~S.}\ \bibnamefont {Dhesi}}, \bibinfo {author}
  {\bibfnamefont {Francesco}\ \bibnamefont {Maccherozzi}}, \bibinfo {author}
  {\bibfnamefont {Vit}\ \bibnamefont {Novak}}, \bibinfo {author} {\bibfnamefont
  {Joerg}\ \bibnamefont {Wunderlich}}, \ and\ \bibinfo {author} {\bibfnamefont
  {Tomas}\ \bibnamefont {Jungwirth}},\ }\bibfield  {title} {\enquote {\bibinfo
  {title} {Current polarity-dependent manipulation of antiferromagnetic
  domains},}\ }\href {\doibase 10.1038/s41565-018-0079-1} {\bibfield  {journal}
  {\bibinfo  {journal} {Nat. Nanotechnol.}\ }\textbf {\bibinfo {volume} {13}},\
  \bibinfo {pages} {362--365} (\bibinfo {year} {2018})}\BibitemShut {NoStop}%
\bibitem [{\citenamefont {Godinho}\ \emph {et~al.}(2018)\citenamefont
  {Godinho}, \citenamefont {Reichlov{\'a}}, \citenamefont {Kriegner},
  \citenamefont {Nov{\'a}k}, \citenamefont {Olejn{\'\i}k}, \citenamefont
  {Ka{\v{s}}par}, \citenamefont {{\v{S}}ob{\'a}{\v{n}}}, \citenamefont
  {Wadley}, \citenamefont {Campion}, \citenamefont {Otxoa} \emph
  {et~al.}}]{godinho_electrically_2018}%
  \BibitemOpen
  \bibfield  {author} {\bibinfo {author} {\bibfnamefont {J}~\bibnamefont
  {Godinho}}, \bibinfo {author} {\bibfnamefont {H}~\bibnamefont
  {Reichlov{\'a}}}, \bibinfo {author} {\bibfnamefont {D}~\bibnamefont
  {Kriegner}}, \bibinfo {author} {\bibfnamefont {V}~\bibnamefont {Nov{\'a}k}},
  \bibinfo {author} {\bibfnamefont {K}~\bibnamefont {Olejn{\'\i}k}}, \bibinfo
  {author} {\bibfnamefont {Z}~\bibnamefont {Ka{\v{s}}par}}, \bibinfo {author}
  {\bibfnamefont {Z}~\bibnamefont {{\v{S}}ob{\'a}{\v{n}}}}, \bibinfo {author}
  {\bibfnamefont {P}~\bibnamefont {Wadley}}, \bibinfo {author} {\bibfnamefont
  {RP}~\bibnamefont {Campion}}, \bibinfo {author} {\bibfnamefont
  {RM}~\bibnamefont {Otxoa}},  \emph {et~al.},\ }\bibfield  {title} {\enquote
  {\bibinfo {title} {Electrically induced and detected {N\'eel} vector reversal
  in a collinear antiferromagnet},}\ }\href {\doibase
  10.1038/s41467-018-07092-2} {\bibfield  {journal} {\bibinfo  {journal} {Nat.
  Commun.}\ }\textbf {\bibinfo {volume} {9}},\ \bibinfo {pages} {4686}
  (\bibinfo {year} {2018})}\BibitemShut {NoStop}%
\bibitem [{\citenamefont {Wortmann}\ \emph {et~al.}(2001)\citenamefont
  {Wortmann}, \citenamefont {Heinze}, \citenamefont {Kurz}, \citenamefont
  {Bihlmayer},\ and\ \citenamefont {Bl\"ugel}}]{PhysRevLett.86.4132}%
  \BibitemOpen
  \bibfield  {author} {\bibinfo {author} {\bibfnamefont {D.}~\bibnamefont
  {Wortmann}}, \bibinfo {author} {\bibfnamefont {S.}~\bibnamefont {Heinze}},
  \bibinfo {author} {\bibfnamefont {Ph.}\ \bibnamefont {Kurz}}, \bibinfo
  {author} {\bibfnamefont {G.}~\bibnamefont {Bihlmayer}}, \ and\ \bibinfo
  {author} {\bibfnamefont {S.}~\bibnamefont {Bl\"ugel}},\ }\bibfield  {title}
  {\enquote {\bibinfo {title} {Resolving complex atomic-scale spin structures
  by spin-polarized scanning tunneling microscopy},}\ }\href {\doibase
  10.1103/PhysRevLett.86.4132} {\bibfield  {journal} {\bibinfo  {journal}
  {Phys. Rev. Lett.}\ }\textbf {\bibinfo {volume} {86}},\ \bibinfo {pages}
  {4132--4135} (\bibinfo {year} {2001})}\BibitemShut {NoStop}%
\bibitem [{\citenamefont {Grzybowski}\ \emph
  {et~al.}(2017{\natexlab{a}})\citenamefont {Grzybowski}, \citenamefont
  {Wadley}, \citenamefont {Edmonds}, \citenamefont {Beardsley}, \citenamefont
  {Hills}, \citenamefont {Campion}, \citenamefont {Gallagher}, \citenamefont
  {Chauhan}, \citenamefont {Novak}, \citenamefont {Jungwirth}, \citenamefont
  {Maccherozzi},\ and\ \citenamefont {Dhesi}}]{PhysRevLett.118.057701}%
  \BibitemOpen
  \bibfield  {author} {\bibinfo {author} {\bibfnamefont {M.~J.}\ \bibnamefont
  {Grzybowski}}, \bibinfo {author} {\bibfnamefont {P.}~\bibnamefont {Wadley}},
  \bibinfo {author} {\bibfnamefont {K.~W.}\ \bibnamefont {Edmonds}}, \bibinfo
  {author} {\bibfnamefont {R.}~\bibnamefont {Beardsley}}, \bibinfo {author}
  {\bibfnamefont {V.}~\bibnamefont {Hills}}, \bibinfo {author} {\bibfnamefont
  {R.~P.}\ \bibnamefont {Campion}}, \bibinfo {author} {\bibfnamefont {B.~L.}\
  \bibnamefont {Gallagher}}, \bibinfo {author} {\bibfnamefont {J.~S.}\
  \bibnamefont {Chauhan}}, \bibinfo {author} {\bibfnamefont {V.}~\bibnamefont
  {Novak}}, \bibinfo {author} {\bibfnamefont {T.}~\bibnamefont {Jungwirth}},
  \bibinfo {author} {\bibfnamefont {F.}~\bibnamefont {Maccherozzi}}, \ and\
  \bibinfo {author} {\bibfnamefont {S.~S.}\ \bibnamefont {Dhesi}},\ }\bibfield
  {title} {\enquote {\bibinfo {title} {Imaging current-induced switching of
  antiferromagnetic domains in cumnas},}\ }\href {\doibase
  10.1103/PhysRevLett.118.057701} {\bibfield  {journal} {\bibinfo  {journal}
  {Phys. Rev. Lett.}\ }\textbf {\bibinfo {volume} {118}},\ \bibinfo {pages}
  {057701} (\bibinfo {year} {2017}{\natexlab{a}})}\BibitemShut {NoStop}%
\bibitem [{\citenamefont {Saidl}\ \emph {et~al.}(2017)\citenamefont {Saidl},
  \citenamefont {N{\v{e}}mec}, \citenamefont {Wadley}, \citenamefont {Hills},
  \citenamefont {Campion}, \citenamefont {Nov{\'a}k}, \citenamefont {Edmonds},
  \citenamefont {Maccherozzi}, \citenamefont {Dhesi}, \citenamefont {Gallagher}
  \emph {et~al.}}]{saidl2017optical}%
  \BibitemOpen
  \bibfield  {author} {\bibinfo {author} {\bibfnamefont {V}~\bibnamefont
  {Saidl}}, \bibinfo {author} {\bibfnamefont {P}~\bibnamefont {N{\v{e}}mec}},
  \bibinfo {author} {\bibfnamefont {P}~\bibnamefont {Wadley}}, \bibinfo
  {author} {\bibfnamefont {V}~\bibnamefont {Hills}}, \bibinfo {author}
  {\bibfnamefont {RP}~\bibnamefont {Campion}}, \bibinfo {author} {\bibfnamefont
  {V}~\bibnamefont {Nov{\'a}k}}, \bibinfo {author} {\bibfnamefont
  {KW}~\bibnamefont {Edmonds}}, \bibinfo {author} {\bibfnamefont
  {F}~\bibnamefont {Maccherozzi}}, \bibinfo {author} {\bibfnamefont
  {SS}~\bibnamefont {Dhesi}}, \bibinfo {author} {\bibfnamefont
  {BL}~\bibnamefont {Gallagher}},  \emph {et~al.},\ }\bibfield  {title}
  {\enquote {\bibinfo {title} {Optical determination of the n{\'e}el vector in
  a cumnas thin-film antiferromagnet},}\ }\href@noop {} {\bibfield  {journal}
  {\bibinfo  {journal} {Nature Photonics}\ }\textbf {\bibinfo {volume} {11}},\
  \bibinfo {pages} {91--96} (\bibinfo {year} {2017})}\BibitemShut {NoStop}%
\bibitem [{\citenamefont {Sun}\ \emph {et~al.}(2019)\citenamefont {Sun},
  \citenamefont {Yi}, \citenamefont {Song}, \citenamefont {Clark},
  \citenamefont {Huang}, \citenamefont {Shan}, \citenamefont {Wu},
  \citenamefont {Huang}, \citenamefont {Gao}, \citenamefont {Chen} \emph
  {et~al.}}]{sun2019giant}%
  \BibitemOpen
  \bibfield  {author} {\bibinfo {author} {\bibfnamefont {Zeyuan}\ \bibnamefont
  {Sun}}, \bibinfo {author} {\bibfnamefont {Yangfan}\ \bibnamefont {Yi}},
  \bibinfo {author} {\bibfnamefont {Tiancheng}\ \bibnamefont {Song}}, \bibinfo
  {author} {\bibfnamefont {Genevieve}\ \bibnamefont {Clark}}, \bibinfo {author}
  {\bibfnamefont {Bevin}\ \bibnamefont {Huang}}, \bibinfo {author}
  {\bibfnamefont {Yuwei}\ \bibnamefont {Shan}}, \bibinfo {author}
  {\bibfnamefont {Shuang}\ \bibnamefont {Wu}}, \bibinfo {author} {\bibfnamefont
  {Di}~\bibnamefont {Huang}}, \bibinfo {author} {\bibfnamefont {Chunlei}\
  \bibnamefont {Gao}}, \bibinfo {author} {\bibfnamefont {Zhanghai}\
  \bibnamefont {Chen}},  \emph {et~al.},\ }\bibfield  {title} {\enquote
  {\bibinfo {title} {Giant nonreciprocal second-harmonic generation from
  antiferromagnetic bilayer cri3},}\ }\href@noop {} {\bibfield  {journal}
  {\bibinfo  {journal} {Nature}\ }\textbf {\bibinfo {volume} {572}},\ \bibinfo
  {pages} {497--501} (\bibinfo {year} {2019})}\BibitemShut {NoStop}%
\bibitem [{\citenamefont {Ni}\ \emph {et~al.}(2021)\citenamefont {Ni},
  \citenamefont {Haglund}, \citenamefont {Wang}, \citenamefont {Xu},
  \citenamefont {Bernhard}, \citenamefont {Mandrus}, \citenamefont {Qian},
  \citenamefont {Mele}, \citenamefont {Kane},\ and\ \citenamefont
  {Wu}}]{ni2021imaging}%
  \BibitemOpen
  \bibfield  {author} {\bibinfo {author} {\bibfnamefont {Zhuoliang}\
  \bibnamefont {Ni}}, \bibinfo {author} {\bibfnamefont {A.~V.}\ \bibnamefont
  {Haglund}}, \bibinfo {author} {\bibfnamefont {H.}~\bibnamefont {Wang}},
  \bibinfo {author} {\bibfnamefont {B.}~\bibnamefont {Xu}}, \bibinfo {author}
  {\bibfnamefont {C.}~\bibnamefont {Bernhard}}, \bibinfo {author}
  {\bibfnamefont {D.~G.}\ \bibnamefont {Mandrus}}, \bibinfo {author}
  {\bibfnamefont {X.}~\bibnamefont {Qian}}, \bibinfo {author} {\bibfnamefont
  {E.~J.}\ \bibnamefont {Mele}}, \bibinfo {author} {\bibfnamefont {C.~L.}\
  \bibnamefont {Kane}}, \ and\ \bibinfo {author} {\bibfnamefont {Liang}\
  \bibnamefont {Wu}},\ }\bibfield  {title} {\enquote {\bibinfo {title} {Imaging
  the n\'eel vector switching in the monolayer antiferromagnet mnpse3 with
  strain-controlled ising order},}\ }\href {\doibase
  10.1038/s41565-021-00885-5} {\bibfield  {journal} {\bibinfo  {journal} {Nat.
  Nanotechnol.}\ }\textbf {\bibinfo {volume} {16}},\ \bibinfo {pages}
  {782--787} (\bibinfo {year} {2021})}\BibitemShut {NoStop}%
\bibitem [{\citenamefont {Erickson}\ \emph {et~al.}(2023)\citenamefont
  {Erickson}, \citenamefont {Abbas~Shah}, \citenamefont {Mahmood},
  \citenamefont {Fescenko}, \citenamefont {Timalsina}, \citenamefont {Binek},\
  and\ \citenamefont {Laraoui}}]{D2RA06440E}%
  \BibitemOpen
  \bibfield  {author} {\bibinfo {author} {\bibfnamefont {Adam}\ \bibnamefont
  {Erickson}}, \bibinfo {author} {\bibfnamefont {Syed~Qamar}\ \bibnamefont
  {Abbas~Shah}}, \bibinfo {author} {\bibfnamefont {Ather}\ \bibnamefont
  {Mahmood}}, \bibinfo {author} {\bibfnamefont {Ilja}\ \bibnamefont
  {Fescenko}}, \bibinfo {author} {\bibfnamefont {Rupak}\ \bibnamefont
  {Timalsina}}, \bibinfo {author} {\bibfnamefont {Christian}\ \bibnamefont
  {Binek}}, \ and\ \bibinfo {author} {\bibfnamefont {Abdelghani}\ \bibnamefont
  {Laraoui}},\ }\bibfield  {title} {\enquote {\bibinfo {title} {Nanoscale
  imaging of antiferromagnetic domains in epitaxial films of cr2o3 via scanning
  diamond magnetic probe microscopy},}\ }\href {\doibase 10.1039/D2RA06440E}
  {\bibfield  {journal} {\bibinfo  {journal} {RSC Adv.}\ }\textbf {\bibinfo
  {volume} {13}},\ \bibinfo {pages} {178--185} (\bibinfo {year}
  {2023})}\BibitemShut {NoStop}%
\bibitem [{\citenamefont {\ifmmode~\check{Z}\else \v{Z}\fi{}elezn\'y}\ \emph
  {et~al.}(2018)\citenamefont {\ifmmode~\check{Z}\else \v{Z}\fi{}elezn\'y},
  \citenamefont {Wadley}, \citenamefont {Olejn\'{\i}k}, \citenamefont
  {Hoffmann},\ and\ \citenamefont {Ohno}}]{zelezny2018SpinTransport}%
  \BibitemOpen
  \bibfield  {author} {\bibinfo {author} {\bibfnamefont {J.}~\bibnamefont
  {\ifmmode~\check{Z}\else \v{Z}\fi{}elezn\'y}}, \bibinfo {author}
  {\bibfnamefont {P.}~\bibnamefont {Wadley}}, \bibinfo {author} {\bibfnamefont
  {K.}~\bibnamefont {Olejn\'{\i}k}}, \bibinfo {author} {\bibfnamefont
  {A.}~\bibnamefont {Hoffmann}}, \ and\ \bibinfo {author} {\bibfnamefont
  {H.}~\bibnamefont {Ohno}},\ }\bibfield  {title} {\enquote {\bibinfo {title}
  {Spin transport and spin torque in antiferromagnetic devices},}\ }\href
  {\doibase 10.1038/s41567-018-0062-7} {\bibfield  {journal} {\bibinfo
  {journal} {Nat. Phys.}\ }\textbf {\bibinfo {volume} {14}},\ \bibinfo {pages}
  {220--228} (\bibinfo {year} {2018})}\BibitemShut {NoStop}%
\bibitem [{\citenamefont {Wang}\ \emph {et~al.}(2021)\citenamefont {Wang},
  \citenamefont {Gao},\ and\ \citenamefont {Xiao}}]{PhysRevLett.127.277201}%
  \BibitemOpen
  \bibfield  {author} {\bibinfo {author} {\bibfnamefont {Chong}\ \bibnamefont
  {Wang}}, \bibinfo {author} {\bibfnamefont {Yang}\ \bibnamefont {Gao}}, \ and\
  \bibinfo {author} {\bibfnamefont {Di}~\bibnamefont {Xiao}},\ }\bibfield
  {title} {\enquote {\bibinfo {title} {Intrinsic nonlinear hall effect in
  antiferromagnetic tetragonal cumnas},}\ }\href {\doibase
  10.1103/PhysRevLett.127.277201} {\bibfield  {journal} {\bibinfo  {journal}
  {Phys. Rev. Lett.}\ }\textbf {\bibinfo {volume} {127}},\ \bibinfo {pages}
  {277201} (\bibinfo {year} {2021})}\BibitemShut {NoStop}%
\bibitem [{\citenamefont {Liu}\ \emph {et~al.}(2021)\citenamefont {Liu},
  \citenamefont {Zhao}, \citenamefont {Huang}, \citenamefont {Wu},
  \citenamefont {Sheng}, \citenamefont {Xiao},\ and\ \citenamefont
  {Yang}}]{PhysRevLett.127.277202}%
  \BibitemOpen
  \bibfield  {author} {\bibinfo {author} {\bibfnamefont {Huiying}\ \bibnamefont
  {Liu}}, \bibinfo {author} {\bibfnamefont {Jianzhou}\ \bibnamefont {Zhao}},
  \bibinfo {author} {\bibfnamefont {Yue-Xin}\ \bibnamefont {Huang}}, \bibinfo
  {author} {\bibfnamefont {Weikang}\ \bibnamefont {Wu}}, \bibinfo {author}
  {\bibfnamefont {Xian-Lei}\ \bibnamefont {Sheng}}, \bibinfo {author}
  {\bibfnamefont {Cong}\ \bibnamefont {Xiao}}, \ and\ \bibinfo {author}
  {\bibfnamefont {Shengyuan~A.}\ \bibnamefont {Yang}},\ }\bibfield  {title}
  {\enquote {\bibinfo {title} {Intrinsic second-order anomalous hall effect and
  its application in compensated antiferromagnets},}\ }\href {\doibase
  10.1103/PhysRevLett.127.277202} {\bibfield  {journal} {\bibinfo  {journal}
  {Phys. Rev. Lett.}\ }\textbf {\bibinfo {volume} {127}},\ \bibinfo {pages}
  {277202} (\bibinfo {year} {2021})}\BibitemShut {NoStop}%
\bibitem [{\citenamefont {Jungwirth}\ \emph {et~al.}(2018)\citenamefont
  {Jungwirth}, \citenamefont {Sinova}, \citenamefont {Manchon}, \citenamefont
  {Marti}, \citenamefont {Wunderlich},\ and\ \citenamefont
  {Felser}}]{Jungwirth2018mutiple_diraction}%
  \BibitemOpen
  \bibfield  {author} {\bibinfo {author} {\bibfnamefont {T.}~\bibnamefont
  {Jungwirth}}, \bibinfo {author} {\bibfnamefont {J.}~\bibnamefont {Sinova}},
  \bibinfo {author} {\bibfnamefont {A.}~\bibnamefont {Manchon}}, \bibinfo
  {author} {\bibfnamefont {X.}~\bibnamefont {Marti}}, \bibinfo {author}
  {\bibfnamefont {J.}~\bibnamefont {Wunderlich}}, \ and\ \bibinfo {author}
  {\bibfnamefont {C.}~\bibnamefont {Felser}},\ }\bibfield  {title} {\enquote
  {\bibinfo {title} {The multiple directions of antiferromagnetic
  spintronics},}\ }\href {\doibase 10.1038/s41567-018-0063-6} {\bibfield
  {journal} {\bibinfo  {journal} {Nat. Phys.}\ }\textbf {\bibinfo {volume}
  {14}},\ \bibinfo {pages} {200--203} (\bibinfo {year} {2018})}\BibitemShut
  {NoStop}%
\bibitem [{\citenamefont {Aapro}\ \emph {et~al.}(2021)\citenamefont {Aapro},
  \citenamefont {Huda}, \citenamefont {Karthikeyan}, \citenamefont
  {Kezilebieke}, \citenamefont {Ganguli}, \citenamefont {Herrero},
  \citenamefont {Huang}, \citenamefont {Liljeroth},\ and\ \citenamefont
  {Komsa}}]{acsnano.1c05532}%
  \BibitemOpen
  \bibfield  {author} {\bibinfo {author} {\bibfnamefont {Markus}\ \bibnamefont
  {Aapro}}, \bibinfo {author} {\bibfnamefont {Md.~Nurul}\ \bibnamefont {Huda}},
  \bibinfo {author} {\bibfnamefont {Jeyakumar}\ \bibnamefont {Karthikeyan}},
  \bibinfo {author} {\bibfnamefont {Shawulienu}\ \bibnamefont {Kezilebieke}},
  \bibinfo {author} {\bibfnamefont {Somesh~C.}\ \bibnamefont {Ganguli}},
  \bibinfo {author} {\bibfnamefont {Héctor~Gonz\'{a}lez}\ \bibnamefont
  {Herrero}}, \bibinfo {author} {\bibfnamefont {Xin}\ \bibnamefont {Huang}},
  \bibinfo {author} {\bibfnamefont {Peter}\ \bibnamefont {Liljeroth}}, \ and\
  \bibinfo {author} {\bibfnamefont {Hannu-Pekka}\ \bibnamefont {Komsa}},\
  }\bibfield  {title} {\enquote {\bibinfo {title} {Synthesis and properties of
  monolayer mnse with unusual atomic structure and antiferromagnetic
  ordering},}\ }\href {\doibase 10.1021/acsnano.1c05532} {\bibfield  {journal}
  {\bibinfo  {journal} {ACS Nano}\ }\textbf {\bibinfo {volume} {15}},\ \bibinfo
  {pages} {13794--13802} (\bibinfo {year} {2021})}\BibitemShut {NoStop}%
\bibitem [{Note1()}]{Note1}%
  \BibitemOpen
  \bibinfo {note} {\label {fn}See Supplemental Material at
  http://link.aps.org/supplemental/xxx, for more details about the derivation
  of the effective Hamiltonian and the numerical calculation of INH effect,
  which include Refs.~\cite
  {VASP,wannier90,godinho_electrically_2018}}\BibitemShut {NoStop}%
\bibitem [{\citenamefont {Sattar}\ \emph {et~al.}(2022)\citenamefont {Sattar},
  \citenamefont {Islam},\ and\ \citenamefont {Canali}}]{PhysRevB.106.085410}%
  \BibitemOpen
  \bibfield  {author} {\bibinfo {author} {\bibfnamefont {Shahid}\ \bibnamefont
  {Sattar}}, \bibinfo {author} {\bibfnamefont {M.~F.}\ \bibnamefont {Islam}}, \
  and\ \bibinfo {author} {\bibfnamefont {C.~M.}\ \bibnamefont {Canali}},\
  }\bibfield  {title} {\enquote {\bibinfo {title} {Monolayer $\mathrm{Mn}x$ and
  janus $x\mathrm{Mn}y$ ($x,y=\mathrm{S}$, se, te): A family of two-dimensional
  antiferromagnetic semiconductors},}\ }\href {\doibase
  10.1103/PhysRevB.106.085410} {\bibfield  {journal} {\bibinfo  {journal}
  {Phys. Rev. B}\ }\textbf {\bibinfo {volume} {106}},\ \bibinfo {pages}
  {085410} (\bibinfo {year} {2022})}\BibitemShut {NoStop}%
\bibitem [{\citenamefont {Grzybowski}\ \emph
  {et~al.}(2017{\natexlab{b}})\citenamefont {Grzybowski}, \citenamefont
  {Wadley}, \citenamefont {Edmonds}, \citenamefont {Beardsley}, \citenamefont
  {Hills}, \citenamefont {Campion}, \citenamefont {Gallagher}, \citenamefont
  {Chauhan}, \citenamefont {Novak}, \citenamefont {Jungwirth}, \citenamefont
  {Maccherozzi},\ and\ \citenamefont
  {Dhesi}}]{grzybowski-ImagingCurrentInducedSwitching-2017}%
  \BibitemOpen
  \bibfield  {author} {\bibinfo {author} {\bibfnamefont {M.~J.}\ \bibnamefont
  {Grzybowski}}, \bibinfo {author} {\bibfnamefont {P.}~\bibnamefont {Wadley}},
  \bibinfo {author} {\bibfnamefont {K.~W.}\ \bibnamefont {Edmonds}}, \bibinfo
  {author} {\bibfnamefont {R.}~\bibnamefont {Beardsley}}, \bibinfo {author}
  {\bibfnamefont {V.}~\bibnamefont {Hills}}, \bibinfo {author} {\bibfnamefont
  {R.~P.}\ \bibnamefont {Campion}}, \bibinfo {author} {\bibfnamefont {B.~L.}\
  \bibnamefont {Gallagher}}, \bibinfo {author} {\bibfnamefont {J.~S.}\
  \bibnamefont {Chauhan}}, \bibinfo {author} {\bibfnamefont {V.}~\bibnamefont
  {Novak}}, \bibinfo {author} {\bibfnamefont {T.}~\bibnamefont {Jungwirth}},
  \bibinfo {author} {\bibfnamefont {F.}~\bibnamefont {Maccherozzi}}, \ and\
  \bibinfo {author} {\bibfnamefont {S.~S.}\ \bibnamefont {Dhesi}},\ }\bibfield
  {title} {\enquote {\bibinfo {title} {Imaging {{Current}}-{{Induced
  Switching}} of {{Antiferromagnetic Domains}} in {{CuMnAs}}},}\ }\href
  {\doibase 10.1103/PhysRevLett.118.057701} {\bibfield  {journal} {\bibinfo
  {journal} {Phys. Rev. Lett.}\ }\textbf {\bibinfo {volume} {118}},\ \bibinfo
  {pages} {057701} (\bibinfo {year} {2017}{\natexlab{b}})}\BibitemShut
  {NoStop}%
\bibitem [{\citenamefont {Bodnar}\ \emph {et~al.}(2019)\citenamefont {Bodnar},
  \citenamefont {Filianina}, \citenamefont {Bommanaboyena}, \citenamefont
  {Forrest}, \citenamefont {Maccherozzi}, \citenamefont {Sapozhnik},
  \citenamefont {Skourski}, \citenamefont {Kl\"aui},\ and\ \citenamefont
  {Jourdan}}]{PhysRevB.99.140409}%
  \BibitemOpen
  \bibfield  {author} {\bibinfo {author} {\bibfnamefont {S.~Yu.}\ \bibnamefont
  {Bodnar}}, \bibinfo {author} {\bibfnamefont {M.}~\bibnamefont {Filianina}},
  \bibinfo {author} {\bibfnamefont {S.~P.}\ \bibnamefont {Bommanaboyena}},
  \bibinfo {author} {\bibfnamefont {T.}~\bibnamefont {Forrest}}, \bibinfo
  {author} {\bibfnamefont {F.}~\bibnamefont {Maccherozzi}}, \bibinfo {author}
  {\bibfnamefont {A.~A.}\ \bibnamefont {Sapozhnik}}, \bibinfo {author}
  {\bibfnamefont {Y.}~\bibnamefont {Skourski}}, \bibinfo {author}
  {\bibfnamefont {M.}~\bibnamefont {Kl\"aui}}, \ and\ \bibinfo {author}
  {\bibfnamefont {M.}~\bibnamefont {Jourdan}},\ }\bibfield  {title} {\enquote
  {\bibinfo {title} {Imaging of current induced n\'eel vector switching in
  antiferromagnetic ${\mathrm{mn}}_{2}\mathrm{Au}$},}\ }\href {\doibase
  10.1103/PhysRevB.99.140409} {\bibfield  {journal} {\bibinfo  {journal} {Phys.
  Rev. B}\ }\textbf {\bibinfo {volume} {99}},\ \bibinfo {pages} {140409}
  (\bibinfo {year} {2019})}\BibitemShut {NoStop}%
\bibitem [{\citenamefont {Nandy}\ and\ \citenamefont
  {Sodemann}(2019)}]{nandy2019symmetry}%
  \BibitemOpen
  \bibfield  {author} {\bibinfo {author} {\bibfnamefont {Snehasish}\
  \bibnamefont {Nandy}}\ and\ \bibinfo {author} {\bibfnamefont {Inti}\
  \bibnamefont {Sodemann}},\ }\bibfield  {title} {\enquote {\bibinfo {title}
  {Symmetry and quantum kinetics of the nonlinear hall effect},}\ }\href
  {\doibase 10.1103/PhysRevB.100.195117} {\bibfield  {journal} {\bibinfo
  {journal} {Phys. Rev. B}\ }\textbf {\bibinfo {volume} {100}},\ \bibinfo
  {pages} {195117} (\bibinfo {year} {2019})}\BibitemShut {NoStop}%
\bibitem [{\citenamefont {Ma}\ \emph {et~al.}(2019)\citenamefont {Ma},
  \citenamefont {Xu}, \citenamefont {Shen}, \citenamefont {MacNeill},
  \citenamefont {Fatemi}, \citenamefont {Chang}, \citenamefont {Valdivia},
  \citenamefont {Wu}, \citenamefont {Du}, \citenamefont {Hsu}, \citenamefont
  {Fang}, \citenamefont {Gibson}, \citenamefont {Watanabe}, \citenamefont
  {Taniguchi}, \citenamefont {Cava}, \citenamefont {Kaxiras}, \citenamefont
  {Lu}, \citenamefont {Lin}, \citenamefont {Fu}, \citenamefont {Gedik},\ and\
  \citenamefont {{Jarillo-Herrero}}}]{ma2019observation}%
  \BibitemOpen
  \bibfield  {author} {\bibinfo {author} {\bibfnamefont {Qiong}\ \bibnamefont
  {Ma}}, \bibinfo {author} {\bibfnamefont {Su-Yang}\ \bibnamefont {Xu}},
  \bibinfo {author} {\bibfnamefont {Huitao}\ \bibnamefont {Shen}}, \bibinfo
  {author} {\bibfnamefont {David}\ \bibnamefont {MacNeill}}, \bibinfo {author}
  {\bibfnamefont {Valla}\ \bibnamefont {Fatemi}}, \bibinfo {author}
  {\bibfnamefont {Tay-Rong}\ \bibnamefont {Chang}}, \bibinfo {author}
  {\bibfnamefont {Andr{\'e}s M.~Mier}\ \bibnamefont {Valdivia}}, \bibinfo
  {author} {\bibfnamefont {Sanfeng}\ \bibnamefont {Wu}}, \bibinfo {author}
  {\bibfnamefont {Zongzheng}\ \bibnamefont {Du}}, \bibinfo {author}
  {\bibfnamefont {Chuang-Han}\ \bibnamefont {Hsu}}, \bibinfo {author}
  {\bibfnamefont {Shiang}\ \bibnamefont {Fang}}, \bibinfo {author}
  {\bibfnamefont {Quinn~D.}\ \bibnamefont {Gibson}}, \bibinfo {author}
  {\bibfnamefont {Kenji}\ \bibnamefont {Watanabe}}, \bibinfo {author}
  {\bibfnamefont {Takashi}\ \bibnamefont {Taniguchi}}, \bibinfo {author}
  {\bibfnamefont {Robert~J.}\ \bibnamefont {Cava}}, \bibinfo {author}
  {\bibfnamefont {Efthimios}\ \bibnamefont {Kaxiras}}, \bibinfo {author}
  {\bibfnamefont {Hai-Zhou}\ \bibnamefont {Lu}}, \bibinfo {author}
  {\bibfnamefont {Hsin}\ \bibnamefont {Lin}}, \bibinfo {author} {\bibfnamefont
  {Liang}\ \bibnamefont {Fu}}, \bibinfo {author} {\bibfnamefont {Nuh}\
  \bibnamefont {Gedik}}, \ and\ \bibinfo {author} {\bibfnamefont {Pablo}\
  \bibnamefont {{Jarillo-Herrero}}},\ }\bibfield  {title} {\enquote {\bibinfo
  {title} {Observation of the nonlinear {{Hall}} effect under
  time-reversal-symmetric conditions},}\ }\href {\doibase
  10.1038/s41586-018-0807-6} {\bibfield  {journal} {\bibinfo  {journal}
  {Nature}\ }\textbf {\bibinfo {volume} {565}},\ \bibinfo {pages} {337--342}
  (\bibinfo {year} {2019})}\BibitemShut {NoStop}%
\bibitem [{\citenamefont {Gao}\ \emph {et~al.}(2014)\citenamefont {Gao},
  \citenamefont {Yang},\ and\ \citenamefont {Niu}}]{gao_field_2014}%
  \BibitemOpen
  \bibfield  {author} {\bibinfo {author} {\bibfnamefont {Yang}\ \bibnamefont
  {Gao}}, \bibinfo {author} {\bibfnamefont {Shengyuan~A.}\ \bibnamefont
  {Yang}}, \ and\ \bibinfo {author} {\bibfnamefont {Qian}\ \bibnamefont
  {Niu}},\ }\bibfield  {title} {\enquote {\bibinfo {title} {Field {Induced}
  {Positional} {Shift} of {Bloch} {Electrons} and {Its} {Dynamical}
  {Implications}},}\ }\href {\doibase 10.1103/PhysRevLett.112.166601}
  {\bibfield  {journal} {\bibinfo  {journal} {Phys. Rev. Lett.}\ }\textbf
  {\bibinfo {volume} {112}},\ \bibinfo {pages} {166601} (\bibinfo {year}
  {2014})}\BibitemShut {NoStop}%
\bibitem [{\citenamefont {Gao}\ and\ \citenamefont
  {Xiao}(2018)}]{gao_orbital_2018}%
  \BibitemOpen
  \bibfield  {author} {\bibinfo {author} {\bibfnamefont {Yang}\ \bibnamefont
  {Gao}}\ and\ \bibinfo {author} {\bibfnamefont {Di}~\bibnamefont {Xiao}},\
  }\bibfield  {title} {\enquote {\bibinfo {title} {Orbital magnetic quadrupole
  moment and nonlinear anomalous thermoelectric transport},}\ }\href {\doibase
  10.1103/PhysRevB.98.060402} {\bibfield  {journal} {\bibinfo  {journal} {Phys.
  Rev. B}\ }\textbf {\bibinfo {volume} {98}},\ \bibinfo {pages} {060402}
  (\bibinfo {year} {2018})}\BibitemShut {NoStop}%
\bibitem [{\citenamefont {Kang}\ \emph {et~al.}(2019)\citenamefont {Kang},
  \citenamefont {Li}, \citenamefont {Sohn}, \citenamefont {Shan},\ and\
  \citenamefont {Mak}}]{Kang2019_fewWTe2}%
  \BibitemOpen
  \bibfield  {author} {\bibinfo {author} {\bibfnamefont {Kaifei}\ \bibnamefont
  {Kang}}, \bibinfo {author} {\bibfnamefont {Tingxin}\ \bibnamefont {Li}},
  \bibinfo {author} {\bibfnamefont {Egon}\ \bibnamefont {Sohn}}, \bibinfo
  {author} {\bibfnamefont {Jie}\ \bibnamefont {Shan}}, \ and\ \bibinfo {author}
  {\bibfnamefont {Kin~Fai}\ \bibnamefont {Mak}},\ }\bibfield  {title} {\enquote
  {\bibinfo {title} {Nonlinear anomalous hall effect in few-layer wte2},}\
  }\href {\doibase 10.1038/s41563-019-0294-7} {\bibfield  {journal} {\bibinfo
  {journal} {Nat. Mater.}\ }\textbf {\bibinfo {volume} {18}},\ \bibinfo {pages}
  {324--328} (\bibinfo {year} {2019})}\BibitemShut {NoStop}%
\bibitem [{\citenamefont {Wang}\ and\ \citenamefont
  {Qian}(2019)}]{Wanghua2019fewWTe2}%
  \BibitemOpen
  \bibfield  {author} {\bibinfo {author} {\bibfnamefont {Hua}\ \bibnamefont
  {Wang}}\ and\ \bibinfo {author} {\bibfnamefont {Xiaofeng}\ \bibnamefont
  {Qian}},\ }\bibfield  {title} {\enquote {\bibinfo {title} {Ferroelectric
  nonlinear anomalous hall effect in few-layer wte$_2$},}\ }\href {\doibase
  10.1038/s41524-019-0257-1} {\bibfield  {journal} {\bibinfo  {journal} {npj
  Computational Materials}\ }\textbf {\bibinfo {volume} {5}},\ \bibinfo {pages}
  {119} (\bibinfo {year} {2019})}\BibitemShut {NoStop}%
\bibitem [{\citenamefont {Williams}\ \emph {et~al.}(2007)\citenamefont
  {Williams}, \citenamefont {DiCarlo},\ and\ \citenamefont
  {Marcus}}]{science.1144657}%
  \BibitemOpen
  \bibfield  {author} {\bibinfo {author} {\bibfnamefont {J.~R.}\ \bibnamefont
  {Williams}}, \bibinfo {author} {\bibfnamefont {L.}~\bibnamefont {DiCarlo}}, \
  and\ \bibinfo {author} {\bibfnamefont {C.~M.}\ \bibnamefont {Marcus}},\
  }\bibfield  {title} {\enquote {\bibinfo {title} {Quantum hall effect in a
  gate-controlled \textit{p-n} junction of graphene},}\ }\href {\doibase
  10.1126/science.1144657} {\bibfield  {journal} {\bibinfo  {journal}
  {Science}\ }\textbf {\bibinfo {volume} {317}},\ \bibinfo {pages} {638--641}
  (\bibinfo {year} {2007})}\BibitemShut {NoStop}%
\bibitem [{\citenamefont {Chen}\ \emph {et~al.}(2010)\citenamefont {Chen},
  \citenamefont {Qin}, \citenamefont {Yang}, \citenamefont {Liu}, \citenamefont
  {Guan}, \citenamefont {Qu}, \citenamefont {Zhang}, \citenamefont {Shi},
  \citenamefont {Xie}, \citenamefont {Yang}, \citenamefont {Wu}, \citenamefont
  {Li},\ and\ \citenamefont {Lu}}]{PhysRevLett.105.176602}%
  \BibitemOpen
  \bibfield  {author} {\bibinfo {author} {\bibfnamefont {J.}~\bibnamefont
  {Chen}}, \bibinfo {author} {\bibfnamefont {H.~J.}\ \bibnamefont {Qin}},
  \bibinfo {author} {\bibfnamefont {F.}~\bibnamefont {Yang}}, \bibinfo {author}
  {\bibfnamefont {J.}~\bibnamefont {Liu}}, \bibinfo {author} {\bibfnamefont
  {T.}~\bibnamefont {Guan}}, \bibinfo {author} {\bibfnamefont {F.~M.}\
  \bibnamefont {Qu}}, \bibinfo {author} {\bibfnamefont {G.~H.}\ \bibnamefont
  {Zhang}}, \bibinfo {author} {\bibfnamefont {J.~R.}\ \bibnamefont {Shi}},
  \bibinfo {author} {\bibfnamefont {X.~C.}\ \bibnamefont {Xie}}, \bibinfo
  {author} {\bibfnamefont {C.~L.}\ \bibnamefont {Yang}}, \bibinfo {author}
  {\bibfnamefont {K.~H.}\ \bibnamefont {Wu}}, \bibinfo {author} {\bibfnamefont
  {Y.~Q.}\ \bibnamefont {Li}}, \ and\ \bibinfo {author} {\bibfnamefont
  {L.}~\bibnamefont {Lu}},\ }\bibfield  {title} {\enquote {\bibinfo {title}
  {Gate-voltage control of chemical potential and weak antilocalization in
  ${\mathrm{bi}}_{2}{\mathrm{se}}_{3}$},}\ }\href {\doibase
  10.1103/PhysRevLett.105.176602} {\bibfield  {journal} {\bibinfo  {journal}
  {Phys. Rev. Lett.}\ }\textbf {\bibinfo {volume} {105}},\ \bibinfo {pages}
  {176602} (\bibinfo {year} {2010})}\BibitemShut {NoStop}%
\bibitem [{\citenamefont {Efetov}\ and\ \citenamefont
  {Kim}(2010)}]{PhysRevLett.105.256805}%
  \BibitemOpen
  \bibfield  {author} {\bibinfo {author} {\bibfnamefont {Dmitri~K.}\
  \bibnamefont {Efetov}}\ and\ \bibinfo {author} {\bibfnamefont {Philip}\
  \bibnamefont {Kim}},\ }\bibfield  {title} {\enquote {\bibinfo {title}
  {Controlling electron-phonon interactions in graphene at ultrahigh carrier
  densities},}\ }\href {\doibase 10.1103/PhysRevLett.105.256805} {\bibfield
  {journal} {\bibinfo  {journal} {Phys. Rev. Lett.}\ }\textbf {\bibinfo
  {volume} {105}},\ \bibinfo {pages} {256805} (\bibinfo {year}
  {2010})}\BibitemShut {NoStop}%
\bibitem [{\citenamefont {Velasco}\ \emph {et~al.}(2016)\citenamefont
  {Velasco}, \citenamefont {Ju}, \citenamefont {Wong}, \citenamefont {Kahn},
  \citenamefont {Lee}, \citenamefont {Tsai}, \citenamefont {Germany},
  \citenamefont {Wickenburg}, \citenamefont {Lu}, \citenamefont {Taniguchi},
  \citenamefont {Watanabe}, \citenamefont {Zettl}, \citenamefont {Wang},\ and\
  \citenamefont {Crommie}}]{nanolett.5b04441}%
  \BibitemOpen
  \bibfield  {author} {\bibinfo {author} {\bibfnamefont {Jairo~Jr.}\
  \bibnamefont {Velasco}}, \bibinfo {author} {\bibfnamefont {Long}\
  \bibnamefont {Ju}}, \bibinfo {author} {\bibfnamefont {Dillon}\ \bibnamefont
  {Wong}}, \bibinfo {author} {\bibfnamefont {Salman}\ \bibnamefont {Kahn}},
  \bibinfo {author} {\bibfnamefont {Juwon}\ \bibnamefont {Lee}}, \bibinfo
  {author} {\bibfnamefont {Hsin-Zon}\ \bibnamefont {Tsai}}, \bibinfo {author}
  {\bibfnamefont {Chad}\ \bibnamefont {Germany}}, \bibinfo {author}
  {\bibfnamefont {Sebastian}\ \bibnamefont {Wickenburg}}, \bibinfo {author}
  {\bibfnamefont {Jiong}\ \bibnamefont {Lu}}, \bibinfo {author} {\bibfnamefont
  {Takashi}\ \bibnamefont {Taniguchi}}, \bibinfo {author} {\bibfnamefont
  {Kenji}\ \bibnamefont {Watanabe}}, \bibinfo {author} {\bibfnamefont {Alex}\
  \bibnamefont {Zettl}}, \bibinfo {author} {\bibfnamefont {Feng}\ \bibnamefont
  {Wang}}, \ and\ \bibinfo {author} {\bibfnamefont {Michael~F.}\ \bibnamefont
  {Crommie}},\ }\bibfield  {title} {\enquote {\bibinfo {title} {Nanoscale
  control of rewriteable doping patterns in pristine graphene/boron nitride
  heterostructures},}\ }\href {\doibase 10.1021/acs.nanolett.5b04441}
  {\bibfield  {journal} {\bibinfo  {journal} {Nano Lett.}\ }\textbf {\bibinfo
  {volume} {16}},\ \bibinfo {pages} {1620--1625} (\bibinfo {year}
  {2016})}\BibitemShut {NoStop}%
\bibitem [{\citenamefont {Shi}\ \emph {et~al.}(2020)\citenamefont {Shi},
  \citenamefont {Kahn}, \citenamefont {Jiang}, \citenamefont {Wang},
  \citenamefont {Tsai}, \citenamefont {Wong}, \citenamefont {Taniguchi},
  \citenamefont {Watanabe}, \citenamefont {Wang}, \citenamefont {Crommie},\
  and\ \citenamefont {Zettl}}]{Shi2020ElectronBeam}%
  \BibitemOpen
  \bibfield  {author} {\bibinfo {author} {\bibfnamefont {Wu}~\bibnamefont
  {Shi}}, \bibinfo {author} {\bibfnamefont {Salman}\ \bibnamefont {Kahn}},
  \bibinfo {author} {\bibfnamefont {Lili}\ \bibnamefont {Jiang}}, \bibinfo
  {author} {\bibfnamefont {Sheng-Yu}\ \bibnamefont {Wang}}, \bibinfo {author}
  {\bibfnamefont {Hsin-Zon}\ \bibnamefont {Tsai}}, \bibinfo {author}
  {\bibfnamefont {Dillon}\ \bibnamefont {Wong}}, \bibinfo {author}
  {\bibfnamefont {Takashi}\ \bibnamefont {Taniguchi}}, \bibinfo {author}
  {\bibfnamefont {Kenji}\ \bibnamefont {Watanabe}}, \bibinfo {author}
  {\bibfnamefont {Feng}\ \bibnamefont {Wang}}, \bibinfo {author} {\bibfnamefont
  {Michael~F.}\ \bibnamefont {Crommie}}, \ and\ \bibinfo {author}
  {\bibfnamefont {Alex}\ \bibnamefont {Zettl}},\ }\bibfield  {title} {\enquote
  {\bibinfo {title} {Reversible writing of high-mobility and
  high-carrier-density doping patterns in two-dimensional van der waals
  heterostructures},}\ }\href {\doibase 10.1038/s41928-019-0351-x} {\bibfield
  {journal} {\bibinfo  {journal} {Nature Electronics}\ }\textbf {\bibinfo
  {volume} {3}},\ \bibinfo {pages} {99--105} (\bibinfo {year}
  {2020})}\BibitemShut {NoStop}%
\bibitem [{\citenamefont {Choi}\ \emph {et~al.}(2021)\citenamefont {Choi},
  \citenamefont {Lee}, \citenamefont {Ngo}, \citenamefont {Hone},\ and\
  \citenamefont {Yoo}}]{aelm.202100449}%
  \BibitemOpen
  \bibfield  {author} {\bibinfo {author} {\bibfnamefont {Min~Sup}\ \bibnamefont
  {Choi}}, \bibinfo {author} {\bibfnamefont {Myeongjin}\ \bibnamefont {Lee}},
  \bibinfo {author} {\bibfnamefont {Tien~Dat}\ \bibnamefont {Ngo}}, \bibinfo
  {author} {\bibfnamefont {James}\ \bibnamefont {Hone}}, \ and\ \bibinfo
  {author} {\bibfnamefont {Won~Jong}\ \bibnamefont {Yoo}},\ }\bibfield  {title}
  {\enquote {\bibinfo {title} {Chemical dopant-free doping by annealing and
  electron beam irradiation on 2d materials},}\ }\href {\doibase
  https://doi.org/10.1002/aelm.202100449} {\bibfield  {journal} {\bibinfo
  {journal} {Advanced Electronic Materials}\ }\textbf {\bibinfo {volume} {7}},\
  \bibinfo {pages} {2100449} (\bibinfo {year} {2021})}\BibitemShut {NoStop}%
\bibitem [{\citenamefont {Lee}\ \emph {et~al.}(2021)\citenamefont {Lee},
  \citenamefont {Lee}, \citenamefont {Kim}, \citenamefont {Kim}, \citenamefont
  {Kim}, \citenamefont {Huh}, \citenamefont {Lee}, \citenamefont {Park},
  \citenamefont {Jeong}, \citenamefont {Kim},\ and\ \citenamefont
  {Lee}}]{Lee2021RemoteModulation}%
  \BibitemOpen
  \bibfield  {author} {\bibinfo {author} {\bibfnamefont {Donghun}\ \bibnamefont
  {Lee}}, \bibinfo {author} {\bibfnamefont {Jea~Jung}\ \bibnamefont {Lee}},
  \bibinfo {author} {\bibfnamefont {Yoon~Seok}\ \bibnamefont {Kim}}, \bibinfo
  {author} {\bibfnamefont {Yeon~Ho}\ \bibnamefont {Kim}}, \bibinfo {author}
  {\bibfnamefont {Jong~Chan}\ \bibnamefont {Kim}}, \bibinfo {author}
  {\bibfnamefont {Woong}\ \bibnamefont {Huh}}, \bibinfo {author} {\bibfnamefont
  {Jaeho}\ \bibnamefont {Lee}}, \bibinfo {author} {\bibfnamefont {Sungmin}\
  \bibnamefont {Park}}, \bibinfo {author} {\bibfnamefont {Hu~Young}\
  \bibnamefont {Jeong}}, \bibinfo {author} {\bibfnamefont {Young~Duck}\
  \bibnamefont {Kim}}, \ and\ \bibinfo {author} {\bibfnamefont {Chul-Ho}\
  \bibnamefont {Lee}},\ }\bibfield  {title} {\enquote {\bibinfo {title} {Remote
  modulation doping in van der waals heterostructure transistors},}\ }\href
  {\doibase 10.1038/s41928-021-00641-6} {\bibfield  {journal} {\bibinfo
  {journal} {Nature Electronics}\ }\textbf {\bibinfo {volume} {4}},\ \bibinfo
  {pages} {664--670} (\bibinfo {year} {2021})}\BibitemShut {NoStop}%
\bibitem [{\citenamefont {Zhao}\ \emph {et~al.}(2017)\citenamefont {Zhao},
  \citenamefont {Xu}, \citenamefont {Pan}, \citenamefont {Zhou}, \citenamefont
  {Zhou},\ and\ \citenamefont {Chai}}]{Zhao2017doping}%
  \BibitemOpen
  \bibfield  {author} {\bibinfo {author} {\bibfnamefont {Yuda}\ \bibnamefont
  {Zhao}}, \bibinfo {author} {\bibfnamefont {Kang}\ \bibnamefont {Xu}},
  \bibinfo {author} {\bibfnamefont {Feng}\ \bibnamefont {Pan}}, \bibinfo
  {author} {\bibfnamefont {Changjian}\ \bibnamefont {Zhou}}, \bibinfo {author}
  {\bibfnamefont {Feichi}\ \bibnamefont {Zhou}}, \ and\ \bibinfo {author}
  {\bibfnamefont {Yang}\ \bibnamefont {Chai}},\ }\bibfield  {title} {\enquote
  {\bibinfo {title} {Doping, contact and interface engineering of
  two-dimensional layered transition metal dichalcogenides transistors},}\
  }\href {\doibase https://doi.org/10.1002/adfm.201603484} {\bibfield
  {journal} {\bibinfo  {journal} {Adv. Funct. Mater.}\ }\textbf {\bibinfo
  {volume} {27}},\ \bibinfo {pages} {1603484} (\bibinfo {year}
  {2017})}\BibitemShut {NoStop}%
\bibitem [{\citenamefont {Winkler}(2003)}]{winkler2003soc}%
  \BibitemOpen
  \bibfield  {author} {\bibinfo {author} {\bibfnamefont {Roland}\ \bibnamefont
  {Winkler}},\ }\href@noop {} {\emph {\bibinfo {title} {Spin—Orbit Coupling
  Effects in Two-Dimensional Electron and Hole Systems}}},\ Vol.\ \bibinfo
  {volume} {191}\ (\bibinfo  {publisher} {Springer Nature, Switzerland AG},\
  \bibinfo {year} {2003})\BibitemShut {NoStop}%
\bibitem [{\citenamefont {Huang}\ \emph {et~al.}(2013)\citenamefont {Huang},
  \citenamefont {Duan},\ and\ \citenamefont {Liu}}]{huang2013existence}%
  \BibitemOpen
  \bibfield  {author} {\bibinfo {author} {\bibfnamefont {Huaqing}\ \bibnamefont
  {Huang}}, \bibinfo {author} {\bibfnamefont {Wenhui}\ \bibnamefont {Duan}}, \
  and\ \bibinfo {author} {\bibfnamefont {Zhirong}\ \bibnamefont {Liu}},\
  }\bibfield  {title} {\enquote {\bibinfo {title} {The existence/absence of
  dirac cones in graphynes},}\ }\href@noop {} {\bibfield  {journal} {\bibinfo
  {journal} {New J. Phys.}\ }\textbf {\bibinfo {volume} {15}},\ \bibinfo
  {pages} {023004} (\bibinfo {year} {2013})}\BibitemShut {NoStop}%
\bibitem [{\citenamefont {Huang}\ \emph {et~al.}(2016)\citenamefont {Huang},
  \citenamefont {Zhou},\ and\ \citenamefont {Duan}}]{huanghqPtSe2}%
  \BibitemOpen
  \bibfield  {author} {\bibinfo {author} {\bibfnamefont {Huaqing}\ \bibnamefont
  {Huang}}, \bibinfo {author} {\bibfnamefont {Shuyun}\ \bibnamefont {Zhou}}, \
  and\ \bibinfo {author} {\bibfnamefont {Wenhui}\ \bibnamefont {Duan}},\
  }\bibfield  {title} {\enquote {\bibinfo {title} {Type-ii dirac fermions in
  the ${\mathrm{ptse}}_{2}$ class of transition metal dichalcogenides},}\
  }\href {\doibase 10.1103/PhysRevB.94.121117} {\bibfield  {journal} {\bibinfo
  {journal} {Phys. Rev. B}\ }\textbf {\bibinfo {volume} {94}},\ \bibinfo
  {pages} {121117} (\bibinfo {year} {2016})}\BibitemShut {NoStop}%
\bibitem [{\citenamefont {Yan}\ \emph {et~al.}(2017)\citenamefont {Yan},
  \citenamefont {Huang}, \citenamefont {Zhang}, \citenamefont {Wang},
  \citenamefont {Yao}, \citenamefont {Deng}, \citenamefont {Wan}, \citenamefont
  {Zhang}, \citenamefont {Arita}, \citenamefont {Yang}, \citenamefont {Sun},
  \citenamefont {Yao}, \citenamefont {Wu}, \citenamefont {Fan}, \citenamefont
  {Duan},\ and\ \citenamefont {Zhou}}]{PtTe2}%
  \BibitemOpen
  \bibfield  {author} {\bibinfo {author} {\bibfnamefont {Mingzhe}\ \bibnamefont
  {Yan}}, \bibinfo {author} {\bibfnamefont {Huaqing}\ \bibnamefont {Huang}},
  \bibinfo {author} {\bibfnamefont {Kenan}\ \bibnamefont {Zhang}}, \bibinfo
  {author} {\bibfnamefont {Eryin}\ \bibnamefont {Wang}}, \bibinfo {author}
  {\bibfnamefont {Wei}\ \bibnamefont {Yao}}, \bibinfo {author} {\bibfnamefont
  {Ke}~\bibnamefont {Deng}}, \bibinfo {author} {\bibfnamefont {Guoliang}\
  \bibnamefont {Wan}}, \bibinfo {author} {\bibfnamefont {Hongyun}\ \bibnamefont
  {Zhang}}, \bibinfo {author} {\bibfnamefont {Masashi}\ \bibnamefont {Arita}},
  \bibinfo {author} {\bibfnamefont {Haitao}\ \bibnamefont {Yang}}, \bibinfo
  {author} {\bibfnamefont {Zhe}\ \bibnamefont {Sun}}, \bibinfo {author}
  {\bibfnamefont {Hong}\ \bibnamefont {Yao}}, \bibinfo {author} {\bibfnamefont
  {Yang}\ \bibnamefont {Wu}}, \bibinfo {author} {\bibfnamefont {Shoushan}\
  \bibnamefont {Fan}}, \bibinfo {author} {\bibfnamefont {Wenhui}\ \bibnamefont
  {Duan}}, \ and\ \bibinfo {author} {\bibfnamefont {Shuyun}\ \bibnamefont
  {Zhou}},\ }\bibfield  {title} {\enquote {\bibinfo {title} {Lorentz-violating
  type-ii dirac fermions in transition metal dichalcogenide ptte2},}\ }\href
  {\doibase 10.1038/s41467-017-00280-6} {\bibfield  {journal} {\bibinfo
  {journal} {Nat. Commun.}\ }\textbf {\bibinfo {volume} {8}},\ \bibinfo {pages}
  {257} (\bibinfo {year} {2017})}\BibitemShut {NoStop}%
\bibitem [{\citenamefont {Zhou}\ \emph {et~al.}(2018)\citenamefont {Zhou},
  \citenamefont {Zhang}, \citenamefont {Li}, \citenamefont {Chen},
  \citenamefont {Shi}, \citenamefont {Tan}, \citenamefont {Gu}, \citenamefont
  {Saleem}, \citenamefont {Wu}, \citenamefont {Pan},\ and\ \citenamefont
  {Song}}]{PhysRevApplied.9.054028}%
  \BibitemOpen
  \bibfield  {author} {\bibinfo {author} {\bibfnamefont {X.~F.}\ \bibnamefont
  {Zhou}}, \bibinfo {author} {\bibfnamefont {J.}~\bibnamefont {Zhang}},
  \bibinfo {author} {\bibfnamefont {F.}~\bibnamefont {Li}}, \bibinfo {author}
  {\bibfnamefont {X.~Z.}\ \bibnamefont {Chen}}, \bibinfo {author}
  {\bibfnamefont {G.~Y.}\ \bibnamefont {Shi}}, \bibinfo {author} {\bibfnamefont
  {Y.~Z.}\ \bibnamefont {Tan}}, \bibinfo {author} {\bibfnamefont {Y.~D.}\
  \bibnamefont {Gu}}, \bibinfo {author} {\bibfnamefont {M.~S.}\ \bibnamefont
  {Saleem}}, \bibinfo {author} {\bibfnamefont {H.~Q.}\ \bibnamefont {Wu}},
  \bibinfo {author} {\bibfnamefont {F.}~\bibnamefont {Pan}}, \ and\ \bibinfo
  {author} {\bibfnamefont {C.}~\bibnamefont {Song}},\ }\bibfield  {title}
  {\enquote {\bibinfo {title} {Strong orientation-dependent spin-orbit torque
  in thin films of the antiferromagnet ${\mathrm{mn}}_{2}\mathrm{Au}$},}\
  }\href {\doibase 10.1103/PhysRevApplied.9.054028} {\bibfield  {journal}
  {\bibinfo  {journal} {Phys. Rev. Applied}\ }\textbf {\bibinfo {volume} {9}},\
  \bibinfo {pages} {054028} (\bibinfo {year} {2018})}\BibitemShut {NoStop}%
\bibitem [{\citenamefont {Meinert}\ \emph {et~al.}(2018)\citenamefont
  {Meinert}, \citenamefont {Graulich},\ and\ \citenamefont
  {Matalla-Wagner}}]{PhysRevApplied.9.064040}%
  \BibitemOpen
  \bibfield  {author} {\bibinfo {author} {\bibfnamefont {Markus}\ \bibnamefont
  {Meinert}}, \bibinfo {author} {\bibfnamefont {Dominik}\ \bibnamefont
  {Graulich}}, \ and\ \bibinfo {author} {\bibfnamefont {Tristan}\ \bibnamefont
  {Matalla-Wagner}},\ }\bibfield  {title} {\enquote {\bibinfo {title}
  {Electrical switching of antiferromagnetic ${\mathrm{mn}}_{2}\mathrm{Au}$ and
  the role of thermal activation},}\ }\href {\doibase
  10.1103/PhysRevApplied.9.064040} {\bibfield  {journal} {\bibinfo  {journal}
  {Phys. Rev. Applied}\ }\textbf {\bibinfo {volume} {9}},\ \bibinfo {pages}
  {064040} (\bibinfo {year} {2018})}\BibitemShut {NoStop}%
\bibitem [{\citenamefont {Du}\ \emph {et~al.}(2018)\citenamefont {Du},
  \citenamefont {Wang}, \citenamefont {Lu},\ and\ \citenamefont
  {Xie}}]{PhysRevLett.121.266601}%
  \BibitemOpen
  \bibfield  {author} {\bibinfo {author} {\bibfnamefont {Z.~Z.}\ \bibnamefont
  {Du}}, \bibinfo {author} {\bibfnamefont {C.~M.}\ \bibnamefont {Wang}},
  \bibinfo {author} {\bibfnamefont {Hai-Zhou}\ \bibnamefont {Lu}}, \ and\
  \bibinfo {author} {\bibfnamefont {X.~C.}\ \bibnamefont {Xie}},\ }\bibfield
  {title} {\enquote {\bibinfo {title} {Band signatures for strong nonlinear
  hall effect in bilayer ${\mathrm{wte}}_{2}$},}\ }\href {\doibase
  10.1103/PhysRevLett.121.266601} {\bibfield  {journal} {\bibinfo  {journal}
  {Phys. Rev. Lett.}\ }\textbf {\bibinfo {volume} {121}},\ \bibinfo {pages}
  {266601} (\bibinfo {year} {2018})}\BibitemShut {NoStop}%
\bibitem [{\citenamefont {Huang}\ \emph {et~al.}(2022)\citenamefont {Huang},
  \citenamefont {Wu}, \citenamefont {Hu}, \citenamefont {Cai}, \citenamefont
  {Li}, \citenamefont {An}, \citenamefont {Feng}, \citenamefont {Ye},
  \citenamefont {Lin}, \citenamefont {Law},\ and\ \citenamefont
  {Wang}}]{nsr2022_TBWSe2}%
  \BibitemOpen
  \bibfield  {author} {\bibinfo {author} {\bibfnamefont {Meizhen}\ \bibnamefont
  {Huang}}, \bibinfo {author} {\bibfnamefont {Zefei}\ \bibnamefont {Wu}},
  \bibinfo {author} {\bibfnamefont {Jinxin}\ \bibnamefont {Hu}}, \bibinfo
  {author} {\bibfnamefont {Xiangbin}\ \bibnamefont {Cai}}, \bibinfo {author}
  {\bibfnamefont {En}~\bibnamefont {Li}}, \bibinfo {author} {\bibfnamefont
  {Liheng}\ \bibnamefont {An}}, \bibinfo {author} {\bibfnamefont {Xuemeng}\
  \bibnamefont {Feng}}, \bibinfo {author} {\bibfnamefont {Ziqing}\ \bibnamefont
  {Ye}}, \bibinfo {author} {\bibfnamefont {Nian}\ \bibnamefont {Lin}}, \bibinfo
  {author} {\bibfnamefont {Kam~Tuen}\ \bibnamefont {Law}}, \ and\ \bibinfo
  {author} {\bibfnamefont {Ning}\ \bibnamefont {Wang}},\ }\bibfield  {title}
  {\enquote {\bibinfo {title} {{Giant nonlinear Hall effect in twisted bilayer
  WSe2}},}\ }\href {\doibase 10.1093/nsr/nwac232} {\bibfield  {journal}
  {\bibinfo  {journal} {Natl. Sci. Rev.}\ } (\bibinfo {year} {2022}),\
  10.1093/nsr/nwac232},\ \bibinfo {note} {nwac232}\BibitemShut {NoStop}%
\bibitem [{\citenamefont {He}\ and\ \citenamefont
  {Weng}(2021)}]{He2021_TBWTe2}%
  \BibitemOpen
  \bibfield  {author} {\bibinfo {author} {\bibfnamefont {Zhihai}\ \bibnamefont
  {He}}\ and\ \bibinfo {author} {\bibfnamefont {Hongming}\ \bibnamefont
  {Weng}},\ }\bibfield  {title} {\enquote {\bibinfo {title} {Giant nonlinear
  hall effect in twisted bilayer wte2},}\ }\href {\doibase
  10.1038/s41535-021-00403-9} {\bibfield  {journal} {\bibinfo  {journal} {npj
  Quantum Materials}\ }\textbf {\bibinfo {volume} {6}},\ \bibinfo {pages} {101}
  (\bibinfo {year} {2021})}\BibitemShut {NoStop}%
\bibitem [{\citenamefont {Duan}\ \emph {et~al.}(2022)\citenamefont {Duan},
  \citenamefont {Jian}, \citenamefont {Gao}, \citenamefont {Peng},
  \citenamefont {Zhong}, \citenamefont {Feng}, \citenamefont {Mao},\ and\
  \citenamefont {Yao}}]{PhysRevLett.129.186801}%
  \BibitemOpen
  \bibfield  {author} {\bibinfo {author} {\bibfnamefont {Junxi}\ \bibnamefont
  {Duan}}, \bibinfo {author} {\bibfnamefont {Yu}~\bibnamefont {Jian}}, \bibinfo
  {author} {\bibfnamefont {Yang}\ \bibnamefont {Gao}}, \bibinfo {author}
  {\bibfnamefont {Huimin}\ \bibnamefont {Peng}}, \bibinfo {author}
  {\bibfnamefont {Jinrui}\ \bibnamefont {Zhong}}, \bibinfo {author}
  {\bibfnamefont {Qi}~\bibnamefont {Feng}}, \bibinfo {author} {\bibfnamefont
  {Jinhai}\ \bibnamefont {Mao}}, \ and\ \bibinfo {author} {\bibfnamefont
  {Yugui}\ \bibnamefont {Yao}},\ }\bibfield  {title} {\enquote {\bibinfo
  {title} {Giant second-order nonlinear hall effect in twisted bilayer
  graphene},}\ }\href {\doibase 10.1103/PhysRevLett.129.186801} {\bibfield
  {journal} {\bibinfo  {journal} {Phys. Rev. Lett.}\ }\textbf {\bibinfo
  {volume} {129}},\ \bibinfo {pages} {186801} (\bibinfo {year}
  {2022})}\BibitemShut {NoStop}%
\bibitem [{\citenamefont {He}\ \emph {et~al.}(2022)\citenamefont {He},
  \citenamefont {Koon}, \citenamefont {Isobe}, \citenamefont {Tan},
  \citenamefont {Hu}, \citenamefont {Neto}, \citenamefont {Fu},\ and\
  \citenamefont {Yang}}]{He2022_grapheneMoire}%
  \BibitemOpen
  \bibfield  {author} {\bibinfo {author} {\bibfnamefont {Pan}\ \bibnamefont
  {He}}, \bibinfo {author} {\bibfnamefont {Gavin Kok~Wai}\ \bibnamefont
  {Koon}}, \bibinfo {author} {\bibfnamefont {Hiroki}\ \bibnamefont {Isobe}},
  \bibinfo {author} {\bibfnamefont {Jun~You}\ \bibnamefont {Tan}}, \bibinfo
  {author} {\bibfnamefont {Junxiong}\ \bibnamefont {Hu}}, \bibinfo {author}
  {\bibfnamefont {Antonio H.~Castro}\ \bibnamefont {Neto}}, \bibinfo {author}
  {\bibfnamefont {Liang}\ \bibnamefont {Fu}}, \ and\ \bibinfo {author}
  {\bibfnamefont {Hyunsoo}\ \bibnamefont {Yang}},\ }\bibfield  {title}
  {\enquote {\bibinfo {title} {Graphene moir\'e superlattices with giant
  quantum nonlinearity of chiral bloch electrons},}\ }\href {\doibase
  10.1038/s41565-021-01060-6} {\bibfield  {journal} {\bibinfo  {journal} {Nat.
  Nanotechnol.}\ }\textbf {\bibinfo {volume} {17}},\ \bibinfo {pages}
  {378--383} (\bibinfo {year} {2022})}\BibitemShut {NoStop}%
\bibitem [{\citenamefont {Zhang}\ \emph {et~al.}(2022)\citenamefont {Zhang},
  \citenamefont {Xiao}, \citenamefont {Zhou}, \citenamefont {Hu}, \citenamefont
  {Xie}, \citenamefont {Yan},\ and\ \citenamefont
  {Law}}]{PhysRevB.106.L041111}%
  \BibitemOpen
  \bibfield  {author} {\bibinfo {author} {\bibfnamefont {Cheng-Ping}\
  \bibnamefont {Zhang}}, \bibinfo {author} {\bibfnamefont {Jiewen}\
  \bibnamefont {Xiao}}, \bibinfo {author} {\bibfnamefont {Benjamin~T.}\
  \bibnamefont {Zhou}}, \bibinfo {author} {\bibfnamefont {Jin-Xin}\
  \bibnamefont {Hu}}, \bibinfo {author} {\bibfnamefont {Ying-Ming}\
  \bibnamefont {Xie}}, \bibinfo {author} {\bibfnamefont {Binghai}\ \bibnamefont
  {Yan}}, \ and\ \bibinfo {author} {\bibfnamefont {K.~T.}\ \bibnamefont
  {Law}},\ }\bibfield  {title} {\enquote {\bibinfo {title} {Giant nonlinear
  hall effect in strained twisted bilayer graphene},}\ }\href {\doibase
  10.1103/PhysRevB.106.L041111} {\bibfield  {journal} {\bibinfo  {journal}
  {Phys. Rev. B}\ }\textbf {\bibinfo {volume} {106}},\ \bibinfo {pages}
  {L041111} (\bibinfo {year} {2022})}\BibitemShut {NoStop}%
\bibitem [{\citenamefont {Chittari}\ \emph {et~al.}(2016)\citenamefont
  {Chittari}, \citenamefont {Park}, \citenamefont {Lee}, \citenamefont {Han},
  \citenamefont {MacDonald}, \citenamefont {Hwang},\ and\ \citenamefont
  {Jung}}]{PhysRevB.94.184428}%
  \BibitemOpen
  \bibfield  {author} {\bibinfo {author} {\bibfnamefont {Bheema~Lingam}\
  \bibnamefont {Chittari}}, \bibinfo {author} {\bibfnamefont {Youngju}\
  \bibnamefont {Park}}, \bibinfo {author} {\bibfnamefont {Dongkyu}\
  \bibnamefont {Lee}}, \bibinfo {author} {\bibfnamefont {Moonsup}\ \bibnamefont
  {Han}}, \bibinfo {author} {\bibfnamefont {Allan~H.}\ \bibnamefont
  {MacDonald}}, \bibinfo {author} {\bibfnamefont {Euyheon}\ \bibnamefont
  {Hwang}}, \ and\ \bibinfo {author} {\bibfnamefont {Jeil}\ \bibnamefont
  {Jung}},\ }\bibfield  {title} {\enquote {\bibinfo {title} {Electronic and
  magnetic properties of single-layer $m\mathrm{P}{X}_{3}$ metal phosphorous
  trichalcogenides},}\ }\href {\doibase 10.1103/PhysRevB.94.184428} {\bibfield
  {journal} {\bibinfo  {journal} {Phys. Rev. B}\ }\textbf {\bibinfo {volume}
  {94}},\ \bibinfo {pages} {184428} (\bibinfo {year} {2016})}\BibitemShut
  {NoStop}%
\bibitem [{\citenamefont {Chu}\ \emph {et~al.}(2020)\citenamefont {Chu},
  \citenamefont {Roh}, \citenamefont {Island}, \citenamefont {Li},
  \citenamefont {Lee}, \citenamefont {Chen}, \citenamefont {Park},
  \citenamefont {Young}, \citenamefont {Lee},\ and\ \citenamefont
  {Hsieh}}]{PhysRevLett.124.027601}%
  \BibitemOpen
  \bibfield  {author} {\bibinfo {author} {\bibfnamefont {Hao}\ \bibnamefont
  {Chu}}, \bibinfo {author} {\bibfnamefont {Chang~Jae}\ \bibnamefont {Roh}},
  \bibinfo {author} {\bibfnamefont {Joshua~O.}\ \bibnamefont {Island}},
  \bibinfo {author} {\bibfnamefont {Chen}\ \bibnamefont {Li}}, \bibinfo
  {author} {\bibfnamefont {Sungmin}\ \bibnamefont {Lee}}, \bibinfo {author}
  {\bibfnamefont {Jingjing}\ \bibnamefont {Chen}}, \bibinfo {author}
  {\bibfnamefont {Je-Geun}\ \bibnamefont {Park}}, \bibinfo {author}
  {\bibfnamefont {Andrea~F.}\ \bibnamefont {Young}}, \bibinfo {author}
  {\bibfnamefont {Jong~Seok}\ \bibnamefont {Lee}}, \ and\ \bibinfo {author}
  {\bibfnamefont {David}\ \bibnamefont {Hsieh}},\ }\bibfield  {title} {\enquote
  {\bibinfo {title} {Linear magnetoelectric phase in ultrathin
  ${\mathrm{mnps}}_{3}$ probed by optical second harmonic generation},}\ }\href
  {\doibase 10.1103/PhysRevLett.124.027601} {\bibfield  {journal} {\bibinfo
  {journal} {Phys. Rev. Lett.}\ }\textbf {\bibinfo {volume} {124}},\ \bibinfo
  {pages} {027601} (\bibinfo {year} {2020})}\BibitemShut {NoStop}%
\bibitem [{\citenamefont {Sivadas}\ \emph {et~al.}(2016)\citenamefont
  {Sivadas}, \citenamefont {Okamoto},\ and\ \citenamefont
  {Xiao}}]{PhysRevLett.117.267203}%
  \BibitemOpen
  \bibfield  {author} {\bibinfo {author} {\bibfnamefont {Nikhil}\ \bibnamefont
  {Sivadas}}, \bibinfo {author} {\bibfnamefont {Satoshi}\ \bibnamefont
  {Okamoto}}, \ and\ \bibinfo {author} {\bibfnamefont {Di}~\bibnamefont
  {Xiao}},\ }\bibfield  {title} {\enquote {\bibinfo {title} {Gate-controllable
  magneto-optic kerr effect in layered collinear antiferromagnets},}\ }\href
  {\doibase 10.1103/PhysRevLett.117.267203} {\bibfield  {journal} {\bibinfo
  {journal} {Phys. Rev. Lett.}\ }\textbf {\bibinfo {volume} {117}},\ \bibinfo
  {pages} {267203} (\bibinfo {year} {2016})}\BibitemShut {NoStop}%
\bibitem [{\citenamefont {Li}\ \emph {et~al.}(2019)\citenamefont {Li},
  \citenamefont {Liu}, \citenamefont {Yu}, \citenamefont {Jiao}, \citenamefont
  {Guan}, \citenamefont {Sheng}, \citenamefont {Yao},\ and\ \citenamefont
  {Yang}}]{PhysRevB.100.205102}%
  \BibitemOpen
  \bibfield  {author} {\bibinfo {author} {\bibfnamefont {Si}~\bibnamefont
  {Li}}, \bibinfo {author} {\bibfnamefont {Ying}\ \bibnamefont {Liu}}, \bibinfo
  {author} {\bibfnamefont {Zhi-Ming}\ \bibnamefont {Yu}}, \bibinfo {author}
  {\bibfnamefont {Yalong}\ \bibnamefont {Jiao}}, \bibinfo {author}
  {\bibfnamefont {Shan}\ \bibnamefont {Guan}}, \bibinfo {author} {\bibfnamefont
  {Xian-Lei}\ \bibnamefont {Sheng}}, \bibinfo {author} {\bibfnamefont {Yugui}\
  \bibnamefont {Yao}}, \ and\ \bibinfo {author} {\bibfnamefont {Shengyuan~A.}\
  \bibnamefont {Yang}},\ }\bibfield  {title} {\enquote {\bibinfo {title}
  {Two-dimensional antiferromagnetic dirac fermions in monolayer
  ${\mathrm{tacote}}_{2}$},}\ }\href {\doibase 10.1103/PhysRevB.100.205102}
  {\bibfield  {journal} {\bibinfo  {journal} {Phys. Rev. B}\ }\textbf {\bibinfo
  {volume} {100}},\ \bibinfo {pages} {205102} (\bibinfo {year}
  {2019})}\BibitemShut {NoStop}%
\bibitem [{\citenamefont {Zhao}\ \emph {et~al.}(2022)\citenamefont {Zhao},
  \citenamefont {Liu}, \citenamefont {Wang}, \citenamefont {Yang},
  \citenamefont {Bellaiche},\ and\ \citenamefont
  {Ma}}]{PhysRevLett.129.187602}%
  \BibitemOpen
  \bibfield  {author} {\bibinfo {author} {\bibfnamefont {Hong~Jian}\
  \bibnamefont {Zhao}}, \bibinfo {author} {\bibfnamefont {Xinran}\ \bibnamefont
  {Liu}}, \bibinfo {author} {\bibfnamefont {Yanchao}\ \bibnamefont {Wang}},
  \bibinfo {author} {\bibfnamefont {Yurong}\ \bibnamefont {Yang}}, \bibinfo
  {author} {\bibfnamefont {Laurent}\ \bibnamefont {Bellaiche}}, \ and\ \bibinfo
  {author} {\bibfnamefont {Yanming}\ \bibnamefont {Ma}},\ }\bibfield  {title}
  {\enquote {\bibinfo {title} {Zeeman effect in centrosymmetric
  antiferromagnetic semiconductors controlled by an electric field},}\ }\href
  {\doibase 10.1103/PhysRevLett.129.187602} {\bibfield  {journal} {\bibinfo
  {journal} {Phys. Rev. Lett.}\ }\textbf {\bibinfo {volume} {129}},\ \bibinfo
  {pages} {187602} (\bibinfo {year} {2022})}\BibitemShut {NoStop}%
\bibitem [{\citenamefont {Yang}\ \emph {et~al.}(2021)\citenamefont {Yang},
  \citenamefont {Zhou}, \citenamefont {Feng},\ and\ \citenamefont
  {Yao}}]{PhysRevB.104.104427}%
  \BibitemOpen
  \bibfield  {author} {\bibinfo {author} {\bibfnamefont {Xiuxian}\ \bibnamefont
  {Yang}}, \bibinfo {author} {\bibfnamefont {Xiaodong}\ \bibnamefont {Zhou}},
  \bibinfo {author} {\bibfnamefont {Wanxiang}\ \bibnamefont {Feng}}, \ and\
  \bibinfo {author} {\bibfnamefont {Yugui}\ \bibnamefont {Yao}},\ }\bibfield
  {title} {\enquote {\bibinfo {title} {Strong magneto-optical effect and
  anomalous transport in the two-dimensional van der waals magnets
  ${\mathrm{fe}}_{n}{\mathrm{gete}}_{2}$ ($n=3$, 4, 5)},}\ }\href {\doibase
  10.1103/PhysRevB.104.104427} {\bibfield  {journal} {\bibinfo  {journal}
  {Phys. Rev. B}\ }\textbf {\bibinfo {volume} {104}},\ \bibinfo {pages}
  {104427} (\bibinfo {year} {2021})}\BibitemShut {NoStop}%
\bibitem [{\citenamefont {Cai}\ \emph {et~al.}(2019)\citenamefont {Cai},
  \citenamefont {Song}, \citenamefont {Wilson}, \citenamefont {Clark},
  \citenamefont {He}, \citenamefont {Zhang}, \citenamefont {Taniguchi},
  \citenamefont {Watanabe}, \citenamefont {Yao}, \citenamefont {Xiao},
  \citenamefont {McGuire}, \citenamefont {Cobden},\ and\ \citenamefont
  {Xu}}]{nanolett.9b01317}%
  \BibitemOpen
  \bibfield  {author} {\bibinfo {author} {\bibfnamefont {Xinghan}\ \bibnamefont
  {Cai}}, \bibinfo {author} {\bibfnamefont {Tiancheng}\ \bibnamefont {Song}},
  \bibinfo {author} {\bibfnamefont {Nathan~P.}\ \bibnamefont {Wilson}},
  \bibinfo {author} {\bibfnamefont {Genevieve}\ \bibnamefont {Clark}}, \bibinfo
  {author} {\bibfnamefont {Minhao}\ \bibnamefont {He}}, \bibinfo {author}
  {\bibfnamefont {Xiaoou}\ \bibnamefont {Zhang}}, \bibinfo {author}
  {\bibfnamefont {Takashi}\ \bibnamefont {Taniguchi}}, \bibinfo {author}
  {\bibfnamefont {Kenji}\ \bibnamefont {Watanabe}}, \bibinfo {author}
  {\bibfnamefont {Wang}\ \bibnamefont {Yao}}, \bibinfo {author} {\bibfnamefont
  {Di}~\bibnamefont {Xiao}}, \bibinfo {author} {\bibfnamefont {Michael~A.}\
  \bibnamefont {McGuire}}, \bibinfo {author} {\bibfnamefont {David~H.}\
  \bibnamefont {Cobden}}, \ and\ \bibinfo {author} {\bibfnamefont {Xiaodong}\
  \bibnamefont {Xu}},\ }\bibfield  {title} {\enquote {\bibinfo {title}
  {Atomically thin crcl3: An in-plane layered antiferromagnetic insulator},}\
  }\href {\doibase 10.1021/acs.nanolett.9b01317} {\bibfield  {journal}
  {\bibinfo  {journal} {Nano Lett.}\ }\textbf {\bibinfo {volume} {19}},\
  \bibinfo {pages} {3993--3998} (\bibinfo {year} {2019})}\BibitemShut {NoStop}%
\bibitem [{\citenamefont {Kresse}\ and\ \citenamefont
  {Furthm\"{u}ller}(1996)}]{VASP}%
  \BibitemOpen
  \bibfield  {author} {\bibinfo {author} {\bibfnamefont {G.}~\bibnamefont
  {Kresse}}\ and\ \bibinfo {author} {\bibfnamefont {J.}~\bibnamefont
  {Furthm\"{u}ller}},\ }\bibfield  {title} {\enquote {\bibinfo {title}
  {Efficiency of ab-initio total energy calculations for metals and
  semiconductors using a plane-wave basis set},}\ }\href@noop {} {\bibfield
  {journal} {\bibinfo  {journal} {Comput. Mater. Sci.}\ }\textbf {\bibinfo
  {volume} {6}},\ \bibinfo {pages} {15} (\bibinfo {year} {1996})}\BibitemShut
  {NoStop}%
\bibitem [{\citenamefont {Mostofi}\ \emph {et~al.}(2008)\citenamefont
  {Mostofi}, \citenamefont {Yates}, \citenamefont {Lee}, \citenamefont {Souza},
  \citenamefont {Vanderbilt},\ and\ \citenamefont {Marzari}}]{wannier90}%
  \BibitemOpen
  \bibfield  {author} {\bibinfo {author} {\bibfnamefont {Arash~A}\ \bibnamefont
  {Mostofi}}, \bibinfo {author} {\bibfnamefont {Jonathan~R}\ \bibnamefont
  {Yates}}, \bibinfo {author} {\bibfnamefont {Young-Su}\ \bibnamefont {Lee}},
  \bibinfo {author} {\bibfnamefont {Ivo}\ \bibnamefont {Souza}}, \bibinfo
  {author} {\bibfnamefont {David}\ \bibnamefont {Vanderbilt}}, \ and\ \bibinfo
  {author} {\bibfnamefont {Nicola}\ \bibnamefont {Marzari}},\ }\bibfield
  {title} {\enquote {\bibinfo {title} {wannier90: A tool for obtaining
  maximally-localised wannier functions},}\ }\href@noop {} {\bibfield
  {journal} {\bibinfo  {journal} {Comput. Phys. Commun.}\ }\textbf {\bibinfo
  {volume} {178}},\ \bibinfo {pages} {685--699} (\bibinfo {year}
  {2008})}\BibitemShut {NoStop}%
\end{thebibliography}%

\end{document}